\documentclass[12pt]{article}
\pdfoutput=1
\usepackage[utf8]{inputenc}
\usepackage[T1]{fontenc}
\usepackage[left=2.55cm, right=2.55cm, top=2.55cm, bottom=2.55cm]{geometry}
\usepackage{amsmath,amssymb,amsbsy}
\usepackage{slashed}
\usepackage{xcolor}
\usepackage{graphicx}
\usepackage{booktabs}
\usepackage{multirow}
\usepackage{hyperref}
\usepackage{float}
\usepackage{parskip}
\usepackage[labelfont=bf]{caption}
\usepackage{cite}

\hypersetup{
    colorlinks=true,
    linkcolor=black,
    citecolor=black,
    urlcolor=black
}

\def\non{\nonumber\\}
\def\gev{{\rm GeV}}

\def\h{\mathrm{hid}}
\newcommand{\Mpl}{M_{\rm Pl}}
\newcommand{\Neff}{\Delta N_{\rm eff}}
\newcommand{\atot}{\alpha_{\rm tot}}
\newcommand{\ah}{\alpha_h}
\newcommand{\Rstar}{R_*}
\newcommand{\Tp}{T_p}
\newcommand{\Tc}{T_c}
\newcommand{\Treh}{T_{\rm reh}}

\newcommand{\nw}[1]{#1}
\newcommand{\nr}[1]{#1}

\newcommand{\lag}{\mathcal{L}}
\newcommand{\mx}{m_A}
\newcommand{\MX}{m_A}
\newcommand{\Phib}{\overline{\Phi}}

\numberwithin{equation}{section}

\begin{document}

\title{\bf\large Natural Supercooling and Reheating along Supersymmetric
Flat Directions and Observable Gravitational Waves at \\
the Einstein Telescope and the Cosmic Explorer}

\author{Jinzheng Li\thanks{\normalsize \href{mailto:li.jinzh@northeastern.edu}{li.jinzh@northeastern.edu}}~ and Pran Nath\thanks{\normalsize \href{mailto:p.nath@northeastern.edu}{p.nath@northeastern.edu}} \\
\textit{\normalsize Department of Physics, Northeastern University, Boston, MA 02115-5000, USA}
}

\date{}

\maketitle

\begin{abstract}

We study supercooled first-order phase transitions in a supersymmetric hidden
sector with a spontaneously broken $U(1)_X$, focusing on the frequency range of
the Einstein Telescope and Cosmic Explorer.  Along the D-flat direction the
tree-level quartic vanishes, so the barrier is generated radiatively by soft
SUSY-breaking splittings.  In the $\overline{\rm DR}$ scheme the gaugino mass
$M_{\tilde\lambda}$ sets the barrier depth, while the soft scalar mass $m_0$
stabilizes the broken vacuum.  For
$M_{\tilde\lambda}/v_X\simeq0.05$--$0.23$, the predicted signal reaches
$\Omega_{\rm GW}h^2\sim3\times10^{-10}$ near the percolation boundary.  The
observable amplitude depends sensitively on the portal coupling $\delta$ through
the hidden-to-visible temperature ratio at percolation: for a cold initial
hidden sector the signal rises from the ET floor at $\delta=10^{-6}$ to
$\Omega_{\rm GW}h^2\simeq7\times10^{-11}$ as the sectors approach thermal contact at
$\delta=10^{-4}$, while a hotter initial hidden sector gives a large signal
already for weak portal coupling.  We follow this evolution with an
11-variable Boltzmann system that separates the cold nucleating exterior from
the reheated true-vacuum interior; reheating mainly enters through the energy
budget and redshift factors.  The same hidden sector can reproduce
$\Omega_{\rm CDM}h^2=0.12$ through relativistic dark-quark freeze-out followed
by entropy dilution from hidden-Higgs decay, with
$m_q\simeq30$--$800\;$keV and
$\Neff\lesssim{\rm few}\times10^{-5}$.

\end{abstract}

\newpage
\section{Introduction}
\label{sec:intro}

Pulsar Timing Array (PTA) observations by
NANOGrav~\cite{NANOGrav:2023gor}, EPTA~\cite{EPTA:2023fyk},
PPTA~\cite{Reardon:2023gzh}, and CPTA~\cite{Xu:2023wog}, building on the
earlier IPTA data combination~\cite{Antoniadis:2022pcn}, have provided
the first evidence for a stochastic gravitational-wave background in the
nano-Hertz band, with sensitivity expected to improve further at the
Square Kilometre Array~\cite{Janssen:2014dka}. A range of cosmological
sources has been examined as possible origins of this
signal~\cite{NANOGrav:2023hvm,EPTA:2023xxk}, among which supercooled
first-order phase transitions (FOPTs) in a hidden
sector~\cite{Schwaller:2015tja,Breitbach:2018ddu,Fairbairn:2019xog,Athron:2023xlk,Athron:2023mer,Kawana:2022lba,Lewicki:2021xku,Madge:2023dxc,Athron:2023rfq,Nakai:2020oit,Ratzinger:2020koh,Feng:2025wvc,Bringmann:2026xcx}
are particularly attractive, since they naturally produce a
gravitational-wave (GW) spectrum with the shape and amplitude suggested
by the data. Because the characteristic frequency of such a signal is
fixed by the temperature of the transition, the same mechanism operating
at higher scales populates the much higher frequency bands accessible to
ground-based interferometers, offering a complementary probe of
hidden-sector dynamics.

In Ref.~\cite{Li:2025nja} we analyzed supercooled FOPTs in a hidden
sector with a spontaneously broken $U(1)_X$ gauge symmetry, and showed
that the mean bubble separation $\Rstar$ must replace the conventional
$\beta_*/H_*$ parametrization for strongly supercooled transitions, and
that the hidden-to-visible temperature ratio $\xi(T_v) = T_h/T_v$ can
shift the GW spectrum by up to four orders of magnitude. That analysis
targeted the PTA band ($f \sim 10^{-9}$--$10^{-7}\;$Hz), corresponding
to a percolation temperature $T_* \equiv \Tp \sim 0.05$--$3\;\gev$. In
the present work we extend this framework to the Einstein Telescope
(ET)~\cite{Punturo:2010zz} band, $f \sim 1$--$10^2\;$Hz, which requires
$T_* \sim 10^4$--$10^8\;\gev$ and correspondingly higher $U(1)_X$
breaking scales $v_X$, and we promote the hidden sector to a
supersymmetric theory with soft breaking communicated by gravity
mediation~\cite{Chamseddine:1982jx,Barbieri:1982eh,Hall:1983iz} (for the
soft-breaking parametrization and reviews
see~
\cite{Nath:1983fp,Nilles:1983ge,Haber:1984rc,Martin:1997ns,Baer:2006rs,Nath:2016qzm}).
Radiative gauge-symmetry breaking driven by soft terms has a long
history in supersymmetric theories~\cite{Nath:2026yma},
with a recent realization given in ~\cite{Akula:2013ioa} and  in a natural MSSM given in Ref.~\cite{Li:2025ypb}.

Supersymmetry offers a natural setting for deep
supercooling~\cite{Hawking:1981fz,Ellis:2018mja,Iso:2017uuu,Randall:2006py}.
Along a D-flat direction the tree-level quartic coupling vanishes
identically by holomorphy of the
superpotential~\cite{Craig:2020jfv,Levi:2022bzt,Witten:1981kv}, so the
potential barrier that drives the transition is generated entirely at
one loop, through the Coleman--Weinberg mechanism acting on the soft
SUSY-breaking mass splittings. Two soft parameters play
complementary roles: the gaugino mass $M_{\tilde\lambda}$ sets the depth
of the radiative barrier, while the soft scalar mass $m_0$ stabilizes
the symmetry-breaking VEV. We work throughout in the
$\overline{\rm DR}$ scheme, in which the boson and fermion contributions to
the hidden-sector effective potential cancel exactly in the supersymmetric
limit, so that the barrier and the resulting supercooling are governed cleanly
by the soft masses. The connection between SUSY flat directions and GW
production was initiated by Craig et al.~\cite{Craig:2020jfv}, who showed
that fixing the mediation scheme ties the GW frequency to the
superpartner spectrum; subsequent work characterized the supercooling
window at weak and strong coupling~\cite{Levi:2022bzt}, developed
model-independent treatments of radiative symmetry
breaking~\cite{Salvio:2023qgb,Salvio:2023ynn}, provided lattice
confirmation~\cite{Dutka:2025oqt}, and examined $U(1)_{B-L}$ breaking in
the SUSY limit~\cite{Haba:2019qol}; closely related are studies of GW
from classically conformal and radiatively broken $U(1)$
extensions~\cite{Jinno:2016knw,Marzola:2017jzl,Okada:2018xdh}.
The present analysis differs from these in treating the phase-transition
scale as a free parameter targeting the ET band, in adopting $\Rstar$
for the strongly supercooled regime, and in scanning
$M_{\tilde\lambda}/v_X$ and $m_0/M_{\tilde\lambda}$ systematically to map
the ET-detectable parameter space.

An essential ingredient is the thermal history of the hidden sector.  The hidden
and visible sectors can communicate through kinetic
mixing~\cite{Holdom:1985ag,Dienes:1996zr,Feldman:2007wj}, and also through
Stueckelberg mass mixing~\cite{Kors:2004ri}; here we focus on kinetic mixing.
The portal coupling $\delta$ controls the energy transfer between the two
baths, while the initial ratio $\xi_0$ encodes the earlier thermal history.
Together they determine the cold hidden-sector temperature at percolation and
therefore the hidden fraction of the total radiation density, which is the main
source of the $\delta$ dependence of the observable GW amplitude.  We compute
this evolution with an extended Boltzmann system whose eleven-component state
vector contains the cold exterior ratio $\xi_f=T_{h,f}/T_v$, the reheated
interior ratio $\xi_t=T_{h,t}/T_v$, five hidden-sector particle yields, and four
nucleation moments.  The nucleation rate is evaluated on the cold branch,
$\Gamma=\Gamma(T_{h,f})$, while latent heat deposited in the true-vacuum
interior affects the volume-averaged energy density, the completion
temperature, and the redshift factors.  As shown in
Section~\ref{sec:coupled_evolution}, increasing $\delta$ can raise
$\xi_{f,p}$ and hence enhance the GW signal, whereas reheating mainly produces a
split between $\xi_f$ and $\xi_t$ and gives a modest redshift correction to the
final spectrum.  Reheating effects in first-order phase transitions have also
been studied in
Refs.~\cite{Rescigno:2025ong,Matuszak:2026xsz,Guo:2020grp,Barman:2026kab}.

The resulting GW signals are large over a broad, untuned region of parameter
space.  Near the percolation boundary we find
$\Omega_{\rm GW}h^2\sim3\times10^{-10}$, nearly three orders of magnitude above
the ET sensitivity floor. The same hidden sector also provides a viable dark matter
candidate.  The dark quark freezes out while relativistic~\cite{Griest:1990kh,Gondolo:1990dk}
and is subsequently diluted by entropy released in the decay of a long-lived
hidden Higgs~\cite{Scherrer:1985zt,Kolb:1990vq,Moroi:1999zb,Giudice:2000ex}.
For the benchmark points considered below this yields
$\Omega_{\rm CDM}h^2=0.12$~\cite{Planck:2018vyg}, with dark quark masses in the
range $m_q\simeq30$--$800\;$keV and with extra radiation at nucleosynthesis far
below current limits, $\Delta N_{\rm eff}\lesssim{\rm few}\times10^{-5}$%
~\cite{Fields:2019pfx,Cyburt:2015mya,Brust:2013ova}.
Connections between dark-sector FOPTs and dark matter have also been explored in
other settings~\cite{Bringmann:2023iuz,Balan:2025uke}.

The paper is organized as follows.  Section~\ref{sec:model} defines the SUSY
$U(1)_X$ model and its effective potential.  Section~\ref{sec:supercooling}
analyzes the radiative supercooling mechanism and presents the
$M_{\tilde\lambda}$ scan and benchmark GW spectra.  In
Section~\ref{sec:coupled_evolution} we develop the coupled thermal evolution,
including the nucleation-moment hierarchy and two-temperature reheating
dynamics, and quantify the effect on the GW spectrum.  Section~\ref{sec:cosmo_constraints}
discusses the dark matter relic density and BBN constraints.  We conclude in
Section~\ref{sec:conclusion}; technical details are collected in the appendices.

\section{The SUSY $U(1)_X$ Model and Effective Potential}
\label{sec:model}

\subsection{Field content and scalar potential}

The hidden sector comprises two chiral superfields $\hat\Phi$, $\hat{\bar\Phi}$
of $U(1)_X$ charge $\pm 1$, a gauge singlet $\hat S$, and a vectorlike pair of
dark quarks $\hat Q$, $\hat{\bar Q}$ of charge $\pm q_X$. The most general
renormalizable superpotential is
\begin{align}
    W_{\h} = \lambda_S \hat{S}\left(\hat{\Phi}\hat{\bar{\Phi}} - v_X^2\right)
    + \mu\,\hat{\Phi}\hat{\bar{\Phi}}
    + y_q \hat{\Phi}\hat{Q}\hat{\bar{Q}},
    \label{eq:superpotential_general}
\end{align}
where $v_X$ is the VEV that breaks the $U(1)_X$ symmetry.
A $\mathbb{Z}_3$ symmetry ($\hat S \to e^{2\pi i/3}\hat S$,
$\hat\Phi \to e^{2\pi i/3}\hat\Phi$,
$\hat{\bar\Phi} \to e^{2\pi i/3}\hat{\bar\Phi}$) forbids the
bare $\mu$-term~\cite{Craig:2020jfv}, giving
\begin{align}
    W_{\h} = \lambda_S \hat{S}\left(\hat{\Phi}\hat{\bar{\Phi}} - v_X^2\right)
    + y_q \hat{\Phi}\hat{Q}\hat{\bar{Q}}.
    \label{eq:superpotential}
\end{align}

The tree-level scalar potential decomposes into $F$-term, $D$-term, and
soft-breaking pieces, $V_0 = V_F + V_D + V_{\rm soft}$, with
$V_F = \sum_i |\partial W_{\h}/\partial\hat\Phi_i|^2$ and the soft terms
generated by gravity-mediated supersymmetry
breaking~\cite{Chamseddine:1982jx,Barbieri:1982eh,Hall:1983iz}:

\begin{align}
    V_F &= |\lambda_S(\Phi\bar\Phi - v_X^2)|^2
    + |\lambda_S S \bar\Phi + y_q Q\bar Q|^2
    + |\lambda_S S \Phi|^2
    + |y_q \Phi \bar Q|^2 + |y_q \Phi Q|^2,
    \label{eq:VF} \\
    V_D &= \frac{g_x^2}{2}\left(|\Phi|^2 - |\bar\Phi|^2
    + q_X|\tilde{q}_L|^2 - q_X|\tilde{q}_R|^2\right)^2,
    \label{eq:VD} \\
    V_{\rm soft} &= m_\Phi^2|\Phi|^2 + m_{\bar\Phi}^2|\bar\Phi|^2
    + m_S^2|S|^2 + m_{\tilde{Q}}^2|\tilde{q}_L|^2
    + m_{\tilde{\bar{Q}}}^2|\tilde{q}_R|^2 \non
    &\quad + \left(A_\lambda \lambda_S S\Phi\bar\Phi
    + A_q y_q \Phi\tilde{q}_L\tilde{q}_R^* + {\rm h.c.}\right)
    + \frac{1}{2}M_{\tilde\lambda}|\tilde\lambda_X|^2,
    \label{eq:Vsoft}
\end{align}
where $M_{\tilde\lambda}$ is the $U(1)_X$ gaugino soft mass and
$m_\Phi^2=m_{\bar\Phi}^2\equiv m_0^2$ the universal soft scalar mass. In the
supersymmetric limit, $F$- and $D$-flatness fix $|\Phi|=|\bar\Phi|=v_X$,
$S=Q=\bar Q=0$, and $V_0=0$; the soft terms lift this flat direction.

Along the $D$-flat direction ($S=\tilde q_{L,R}=0$, $|\Phi|=|\bar\Phi|$) the
$D$-term vanishes identically, and the dynamics reduce to a single canonically
normalized radial field $\phi_c$, defined by $\Phi=\bar\Phi=\phi_c/2$
(Eq.~(\ref{eq:fluct_param}); Appendix~\ref{sec:appendix_potential}), for which
the gauge-boson mass is $m_A^2=g_x^2\phi_c^2$. The symmetry-breaking vacuum
$|\Phi|=|\bar\Phi|=v_X$ corresponds to $\phi_c=2v_X$, equivalently
$\Phi\bar\Phi=\phi_c^2/4=v_X^2$ from $F$-flatness in Eq.~(\ref{eq:VF}). Along
this direction $V_0$ reduces to the single-field tree-level potential
\begin{align}
    V_{\rm tree}(\phi_c) = -\frac{m^2}{2}\phi_c^2
    + \frac{\lambda_h}{4}\phi_c^4\,,
    \qquad
    \lambda_h \equiv \frac{\lambda_S^2}{4},
    \quad
    m^2 \equiv \lambda_S^2 v_X^2 - m_0^2\,,
    \label{eq:Vtree_flat}
\end{align}
where $\lambda_h=\lambda_S^2/4$ is the effective quartic and $m^2$ collects the
$F$-term ($+\lambda_S^2 v_X^2$) and soft-mass ($-m_0^2$) contributions to the
curvature at the origin. The same $m^2$ sets the field-dependent Higgs and
pseudoscalar masses $m_h^2=3\lambda_h\phi_c^2-m^2$ and
$m_{G_h^0}^2=\lambda_h\phi_c^2-m^2$ (Table~\ref{tab:masses}).
In the flat-direction benchmarks below we take the limit $\lambda_S\to0$ at
fixed radiatively generated breaking scale $v_X$.  In this limit the singlet
sector decouples, and $v_X$ should be understood as the renormalization
condition fixing the Coleman--Weinberg minimum, rather than as an $F$-flat VEV
enforced by the $\lambda_S S(\Phi\bar\Phi-v_X^2)$ term.

\subsection{Field-dependent mass spectrum}

\begin{table}[t]
\centering
\small
\begin{tabular}{ll|cc}
\hline
\textbf{Origin} & Particle & DOF $g_i$ & $m_i^2(\phi_c)$ \\
\hline
\multirow{3}{*}{\shortstack[l]{\textit{Massive vector}\\\textit{supermultiplet}}}
& $A_\mu$ (gauge boson) & 3 & $g_x^2\phi_c^2$ \\
& $\tilde\chi_+$ (Majorana eigenstate) & 2 & $m_+^2(\phi_c)$ \\
& $\tilde\chi_-$ (Majorana eigenstate) & 2 & $m_-^2(\phi_c)$ \\
\hline
\multirow{4}{*}{\shortstack[l]{\textit{Scalar sector}\\\textit{of $\Phi,\bar\Phi$}}}
& $\sigma$ (lighter Higgs scalar) & 1 & $g_x^2\phi_c^2 + m_0^2$ \\
& $h$ (Higgs) & 1 & $3\lambda_h\phi_c^2 - m^2$ \\
& $G_h^0$ (pseudoscalar) & 1 & $\lambda_h\phi_c^2 - m^2$ \\
& $G$ (would-be Goldstone) & 1 & $\lambda_h\phi_c^2 - m^2 + \xi_{\rm gauge} g_x^2\phi_c^2$ \\
\hline
\multirow{2}{*}{\shortstack[l]{\textit{Dark quark}\\\textit{sector $\hat Q,\hat{\bar Q}$}}}
& $q$ (Dirac dark quark) & 4 & $y_q^2\phi_c^2/2$ \\
& $\tilde q$ (dark squarks) & 4 & $y_q^2\phi_c^2/2 + m_{\tilde Q}^2$ \\
\hline
\end{tabular}
\caption{Field-dependent mass spectrum along the D-flat direction. The
Majorana eigenstates $\tilde\chi_\pm$ (2~DOF each) diagonalize the
gaugino--Higgsino matrix~(\ref{eq:ferm_mass_matrix}); the soft mass
$m_0^2$ lifts $\sigma$ above the gauge boson, breaking the bosonic
degeneracy. The singlet $\hat{S}$ fields (at $S = 0$) are
omitted as non-contributing.
The would-be Goldstone $G$ is retained in the CW sum ($\mathcal{C}_G=3/2$)
with its general $R_{\xi_{\rm gauge}}$ mass; the numerical analysis uses Landau gauge
($\xi_{\rm gauge}=0$), where $m_G^2=\lambda_h\phi_c^2-m^2$ (degenerate with $G_h^0$) and
the unphysical gauge and ghost determinants vanish
(Appendix~\ref{sec:appendix_potential}). The dark quark sector is negligible (Section~\ref{sec:dq_spectrum}).}
\label{tab:masses}
\end{table}

The complete field-dependent mass spectrum is listed in
Table~\ref{tab:masses}.
Following the parametrization of Appendix~\ref{sec:appendix_potential}, we
expand $\Phi$ and $\bar\Phi$ around the flat-direction background
$\langle\Phi\rangle = \langle\bar\Phi\rangle = \phi_c/2$
as\label{sec:dof_decomposition}
\begin{align}
   \Phi &= \frac{1}{\sqrt2}\!\left[\frac{\phi_c}{\sqrt2}
     +\frac{h+\sigma}{\sqrt2}+i\,\frac{G_h^0+G}{\sqrt2}\right],\nonumber\\
   \bar\Phi &= \frac{1}{\sqrt2}\!\left[\frac{\phi_c}{\sqrt2}
     +\frac{h-\sigma}{\sqrt2}+i\,\frac{G_h^0-G}{\sqrt2}\right],
   \label{eq:fluct_param}
\end{align}
which coincides with the parametrization of
Appendix~\ref{sec:appendix_potential} (which uses the same background field
$\phi_c$ and field names $h$, $\sigma$, $G_h^0$, $G$). Equivalently, the four canonically
normalized real eigenmodes are the symmetric ($\Phi+\bar\Phi$) and
antisymmetric ($\Phi-\bar\Phi$) combinations
\begin{align}
   h &= \mathrm{Re}(\Phi+\bar\Phi)-\phi_c, &
   \sigma &= \mathrm{Re}(\Phi-\bar\Phi),\nonumber\\
   G_h^0 &= \mathrm{Im}(\Phi+\bar\Phi), &
   G &= \mathrm{Im}(\Phi-\bar\Phi)\,.
   \label{eq:rotate_DOF}
\end{align}
Here $h$ is the Higgs scalar, $\sigma$ the lighter Higgs scalar,
$G_h^0$ the physical pseudoscalar, and $G$ the Goldstone boson;
all four are canonically normalised.
Under $\Phi\to e^{ig_x\alpha}\Phi$, $\bar\Phi\to e^{-ig_x\alpha}\bar\Phi$,
only $G$ shifts ($\delta G = g_x\alpha\phi_c$), identifying it as the
would-be Goldstone absorbed by $A_\mu$; the remaining three,
$h$, $\sigma$, and $G_h^0$, are gauge-invariant physical scalars.
The lighter Higgs scalar $\sigma$ parametrises the departure from
D-flatness ($|\Phi|\neq|\bar\Phi|$); its mass $m_\sigma^2 = g_x^2\phi_c^2
+ m_0^2$ receives contributions from the D-term and soft SUSY breaking,
lifting it above $m_A^2 = g_x^2\phi_c^2$ in the SUSY-breaking vacuum.
In the SUSY limit $m_0\to 0$, $m_\sigma^2 = m_A^2$, restoring the
bosonic mass degeneracy of the massive $\mathcal{N}=1$ vector
supermultiplet; $m_0$ breaks this degeneracy, generating the boson--fermion
mass splitting that sources the radiative potential.

In the fermionic sector,\label{sec:fermion_dof} the gaugino $\tilde\lambda_X$
pairs with the antisymmetric Higgsino combination
$\tilde{H}_- = (\tilde\Phi-\tilde{\bar\Phi})/\sqrt{2}$ --- the only combination
that couples to it at the symmetric background
$\langle\Phi\rangle=\langle\bar\Phi\rangle$
(Appendix~\ref{sec:appendix_potential}) --- while the orthogonal combination
$\tilde{H}_+ = (\tilde\Phi+\tilde{\bar\Phi})/\sqrt{2}$ remains massless.
The $2\times2$ Majorana mass matrix in the
$(\tilde\lambda_X,\,\tilde{H}_-)$ basis is
\begin{align}
   \mathcal{M}_F = \begin{pmatrix}
   M_{\tilde\lambda} & g_x\phi_c \\[3pt]
   g_x\phi_c & 0
   \end{pmatrix},
   \label{eq:ferm_mass_matrix}
\end{align}
with eigenvalues
\begin{align}
   m_\pm(\phi_c) = \frac{\sqrt{M_{\tilde\lambda}^2+4g_x^2\phi_c^2}
   \pm M_{\tilde\lambda}}{2}\,,
   \label{eq:mgaugino}
\end{align}
corresponding to two Majorana eigenstates $\tilde\chi_\pm$.
\label{sec:effpot}
The dark quark sector contributes negligibly to the effective potential
($(y_q/g_x)^4 \sim 10^{-38}$), but it does enter the hidden-sector radiation
energy density (Section~\ref{sec:dq_in_PT}).

\subsection{Finite-temperature effective potential}

The first-order phase transition is governed by the finite-temperature
effective potential~\cite{Feng:2024pab,Li:2025nja,Li:2026pjy,Aboubrahim:2022bzk,Li:2023nez,Nath:2024mgr},
given by
\begin{align}
    V_{\rm eff}(\phi_c, T) = V_{\rm tree}(\phi_c) + V_{\rm CW}(\phi_c)
    + \Delta V_T(\phi_c, T) + V_{\rm daisy}(\phi_c, T),
    \label{eq:Veff_full}
\end{align}
where the one-loop Coleman-Weinberg (CW) potential
is~\cite{Coleman:1973jx,Quiros:1999jp}
\begin{align}
    V_{\rm CW}(\phi_c) = \sum_i \frac{(-1)^{2s_i}\,g_i}{64\pi^2}\,
    m_i^4(\phi_c)\left[\ln\frac{m_i^2(\phi_c)}{\mu_R^2}
    - \mathcal{C}_i\right],
    \label{eq:VCW_general}
\end{align}
where the sum runs over the field-dependent spectrum of
Table~\ref{tab:masses} and $\mathcal{C}_i = 3/2$ for all determinants in the
$\overline{\rm DR}$ scheme adopted throughout~\cite{Martin:1997ns}; the scheme
and gauge dependence of the effective potential is reviewed
in~\cite{Patel:2011th}. 
 In Appendix~\ref{sec:appendix_potential} the CW potential in  the $R_{\xi_{\rm gauge}}$
 gauge is derived. 
In the analysis here we used the Landau gauge ($\xi_{\rm gauge}=0$). In this gauge the CW potential
is given by

\begin{align}
V_{\rm CW}(\phi_c) = \frac{1}{64\pi^2}\Big\{\,
& 3\,m_A^4\,L(m_A^2)
+ m_\sigma^4\,L(m_\sigma^2)
+ m_h^4\,L(m_h^2)
+ m_{G_h^0}^4\,L(m_{G_h^0}^2) \nonumber\\
&+ m_G^4\,L(m_G^2)
- 2\,m_+^4\,L(m_+^2)
- 2\,m_-^4\,L(m_-^2)
\,\Big\},
\label{eq:VCW_landau}
\end{align}
where $L(x)\equiv\ln(x/\mu_R^2)-3/2$ and, from Table~\ref{tab:masses}, the
field-dependent masses are $m_A^2=g_x^2\phi_c^2$,
$m_\sigma^2=g_x^2\phi_c^2+m_0^2$, $m_h^2=3\lambda_h\phi_c^2-m^2$,
$m_{G_h^0}^2=m_G^2=\lambda_h\phi_c^2-m^2$, and $m_\pm$ are the
gaugino--Higgsino eigenvalues of Eq.~(\ref{eq:mgaugino}). The finite-temperature correction
\nr{to the potential is given by}~\cite{Dolan:1973qd,Quiros:1999jp}
\begin{align}
    \Delta V_T(\phi_c, T) = \frac{T^4}{2\pi^2}\left[
    \sum_{i\in\text{bosons}} g_i\,J_b\!\left(\frac{m_i^2(\phi_c)}{T^2}
    \right)
    + \sum_{i\in\text{fermions}} g_i\,J_f\!\left(\frac{m_i^2(\phi_c)}{T^2}
    \right)\right],
    \label{eq:VT}
\end{align}
where $J_b(y) = \int_0^\infty dx\,x^2\,\ln(1 - e^{-\sqrt{x^2+y}})$,
$J_f(y) = \int_0^\infty dx\,x^2\,\ln(1 + e^{-\sqrt{x^2+y}})$,
\label{eq:thermal_content}
and the field content is as in Table~\ref{tab:masses}.
\nr{In addition there is contribution from daisy resummation}
~\cite{Arnold:1992rz,Parwani:1991gq,Carrington:1991hz,Curtin:2016urg}:
\begin{align}
    V_{\rm daisy}(\phi_c, T) = -\frac{T}{12\pi}\sum_{i\in\text{bosons}}
    g_i^{\rm ring}\left[\Big(m_i^2(\phi_c) + \Pi_i(T)\Big)^{3/2}
    - \big|m_i^2(\phi_c)\big|^{3/2}\right],
    \label{eq:Vdaisy}
\end{align}
where
$\Pi_A(T) = g_x^2 T^2$ and
$\Pi_\phi(T) = (\lambda_h/4 + g_x^2/2)T^2$ \nr{are the temperature depend Debye mass corrections}. 
\label{eq:PiA}
\label{eq:Piphi}
The ring bosons are $A_\mu^L$, $h$, $G_h^0$, $\sigma$  with one 
degree of freedom each.

\section{Supercooling from Radiative Symmetry Breaking in SUSY Flat Directions}
\label{sec:supercooling}
\label{sec:PT}

\subsection{Radiative symmetry breaking in SUSY flat directions}
\label{sec:lambdaS0}

Along the flat direction the entire potential barrier is generated at one loop
by the Coleman--Weinberg potential, Eq.~(\ref{eq:VCW_landau}). For
$\lambda_h=0$ the scalars $h$, $G_h^0$, and $G$ have the $\phi_c$-independent
mass $m_0^2$ and contribute only a constant shift, so the $\phi_c$-dependent
part of $V_{\rm CW}$---and hence the barrier---comes entirely from the massive
$\mathcal{N}=1$ vector supermultiplet: the gauge boson $A_\mu$ (3 bosonic DOF),
the D-scalar $\sigma$ (1 bosonic DOF), and the two Majorana gaugino--Higgsino
eigenstates $\tilde\chi_\pm$ (2 fermionic DOF each). Dropping the constant, this
contribution is
\begin{align}
V_{\rm CW}^{\rm gauge}(\phi_c) &= \frac{1}{64\pi^2}\left[
    3\,m_A^4\,L(m_A^2)
  + m_\sigma^4\,L(m_\sigma^2)
  - 2\!\sum_{i=\pm}\!m_i^4\,L(m_i^2)
\right],
\label{eq:VCW_gauge_full}
\end{align}
with $L(x)=\ln(x/\mu_R^2)-3/2$, $m_A^2=g_x^2\phi_c^2$,
$m_\sigma^2=g_x^2\phi_c^2+m_0^2$, and $m_\pm$ the eigenvalues of
Eq.~(\ref{eq:mgaugino}).

In the supersymmetric limit ($m_0,M_{\tilde\lambda}\to0$) all of these states
share the common mass $m_A=g_x\phi_c$. With the universal $\overline{\rm DR}$
constant $\mathcal{C}_i=3/2$~\cite{Martin:1997ns}, the four bosonic ($3+1$) and
four fermionic ($2+2$) degrees of freedom then enter
Eq.~(\ref{eq:VCW_gauge_full}) with equal masses and opposite signs, so
$V_{\rm CW}^{\rm gauge}$ vanishes identically and no barrier forms. (The
complete boson--fermion degree-of-freedom counting, including the $R_{\xi_{\rm gauge}}$
Goldstone, unphysical-gauge, and ghost determinants, is given in
Appendix~\ref{sec:appendix_potential}.) Soft breaking lifts this degeneracy in
two complementary ways: the soft scalar mass raises the D-scalar to
$m_\sigma^2=g_x^2\phi_c^2+m_0^2$, a bosonic contribution $\propto m_0^2$ that
stabilizes the VEV, while the gaugino mass $M_{\tilde\lambda}$ splits the
eigenstates $\tilde\chi_\pm$ by $\pm M_{\tilde\lambda}g_x\phi_c$, a compensating
fermionic contribution. The residual boson--fermion mass splitting is the
source of the radiative barrier.

\subsection{Natural supercooling}
\label{sec:supertrace}
\label{sec:scheme_comparison}
\label{sec:vcw}

A first-order transition is \emph{naturally} supercooled---as opposed to merely
thermally supercooled---when the zero-temperature effective potential already
contains a barrier separating the symmetric and broken phases. Writing the
$T=0$ potential as
\begin{equation}
  V_0(\phi_c)=V_{\rm tree}(\phi_c)+V_{\rm CW}(\phi_c),
  \label{eq:V0_zeroT}
\end{equation}
the $U(1)_X$-breaking global minimum lies at $\phi_c=\langle\phi_c\rangle\neq0$,
and a barrier separates it from the origin provided $\phi_c=0$ remains a
\emph{local} minimum at $T=0$. Since $V_0$ depends on $\phi_c$ only through the
field-dependent masses $m_i^2(\phi_c^2)$, it is even in $\phi_c$ and the origin
is automatically a stationary point; the requirement therefore reduces to a
positive curvature there,
\begin{equation}
  \left.\frac{d^2V_0}{d\phi_c^2}\right|_{\phi_c=0}>0 .
  \label{eq:barrier_condition}
\end{equation}
A barrier satisfying Eq.~(\ref{eq:barrier_condition}) persists to arbitrarily
low temperature, so the transition can be delayed far below $T_c$; a purely
thermal barrier, by contrast, generated by the cubic and daisy terms of
$\Delta V_T$, disappears as $T\to0$ and yields only mild supercooling.

This condition is met easily here. At the origin only the two states that
remain massive contribute to the curvature of $V_{\rm CW}$. These are the D-scalar
$\sigma$ ($m_\sigma^2=g_x^2\phi_c^2+m_0^2$) and the heavy gaugino $\tilde\chi_+$
($m_+^2\to M_{\tilde\lambda}^2$), since the gauge boson and the light gaugino
$\tilde\chi_-$ are massless there and the Higgs-sector scalars $h,G_h^0,G$ are
$\phi_c$-independent on the flat direction ($\lambda_h=0$). Adding the tree term,
the curvature is given by  ( see Appendix~\ref{sec:app_barrier} for details)
\begin{equation}
  \left.\frac{d^2V_0}{d\phi_c^2}\right|_{0}=
  m_0^2+\frac{g_x^2}{16\pi^2}\left[
  m_0^2\!\left(\ln\frac{m_0^2}{\mu_R^2}-1\right)
  -4M_{\tilde\lambda}^2\!\left(\ln\frac{M_{\tilde\lambda}^2}{\mu_R^2}-1\right)
  \right].
  \label{eq:V0pp}
\end{equation}
The one-loop term is suppressed by $g_x^2/16\pi^2$ relative to the tree-level
soft mass, so for perturbative coupling
$d^2V_0/d\phi_c^2|_{0}\simeq m_0^2>0$, the soft scalar mass keeps the origin a
local minimum, and the zero-temperature barrier of natural supercooling is
present without any thermal input. This is the qualitative advantage of the
$m^2\neq0$, $\lambda=0$ realization over the classically scale-invariant case
($m^2=0$, $\lambda\neq0$), in which the origin is only marginally stable and a
$T=0$ barrier is not guaranteed.

This radiative breaking is, however, not automatically stable, and demanding
stability constrains the soft spectrum: the gaugino loop, which lowers
$V_{\rm CW}$ at large field, can drive the potential unbounded below, so the
symmetry-breaking vacuum is not guaranteed to be a genuine minimum. Expanding the
full effective potential about the symmetry-breaking point $\phi_0$,
\begin{equation}
  V_{\rm eff}(\phi_c)=V_{\rm eff}(\phi_0)
  +\tfrac12 m_\rho^2(\phi_c-\phi_0)^2
  +\frac{1}{3!}V_{\rm eff}^{(3)}(\phi_0)(\phi_c-\phi_0)^3
  +\frac{1}{4!}V_{\rm eff}^{(4)}(\phi_0)(\phi_c-\phi_0)^4+\cdots,
  \label{eq:Veff_expand}
\end{equation}
the vacuum is stable if and only if $\phi_0$ is a stationary point,
$V_{\rm eff}'(\phi_0)=0$, with positive curvature
$m_\rho^2\equiv V_{\rm eff}''(\phi_0)>0$. Differentiating the one-loop potential
Eq.~(\ref{eq:VCW_landau}) twice gives the exact radial mass
(Appendix~\ref{sec:app_stability})
\begin{equation}
  m_\rho^2 = m_0^2 + \frac{1}{32\pi^2}\sum_i (-1)^{2s_i}g_i\!\left[
  \left(\frac{dm_i^2}{d\phi_c}\right)^{\!2}\ln\frac{m_i^2}{\mu_R^2}
  + m_i^2\,\frac{d^2m_i^2}{d\phi_c^2}\!\left(\ln\frac{m_i^2}{\mu_R^2}-1\right)
  \right]_{\phi_0},
  \label{eq:mrho}
\end{equation}
where $m_0^2=V_{\rm tree}''$ is the tree-level contribution and the sum runs over
the spectrum of Table~\ref{tab:masses} and  
the $\phi_c$-independent scalars, with
$dm_i^2/d\phi_c=0$, drop out. The constraint $m_\rho^2>0$ is imposed 
throughout the parameter scan, and  is satisfied for the soft hierarchy
$m_0\simeq(2.75$--$3.25)\,M_{\tilde\lambda}$ of the benchmarks
(Section~\ref{sec:Mhalf_scan}), where the positive D-scalar contribution
(proportional to $m_0^2$) outweighs the negative gaugino one
(proportional to $M_{\tilde\lambda}^2$), so the D-scalar soft mass stabilizes the vacuum
on its own.

\subsection{Parameter space and benchmark GW spectra}
\label{sec:Mhalf_scan}
\label{sec:benchmarks_xi1}
\begin{figure}[tbp]
  \centering
  \includegraphics[width=0.85\linewidth]{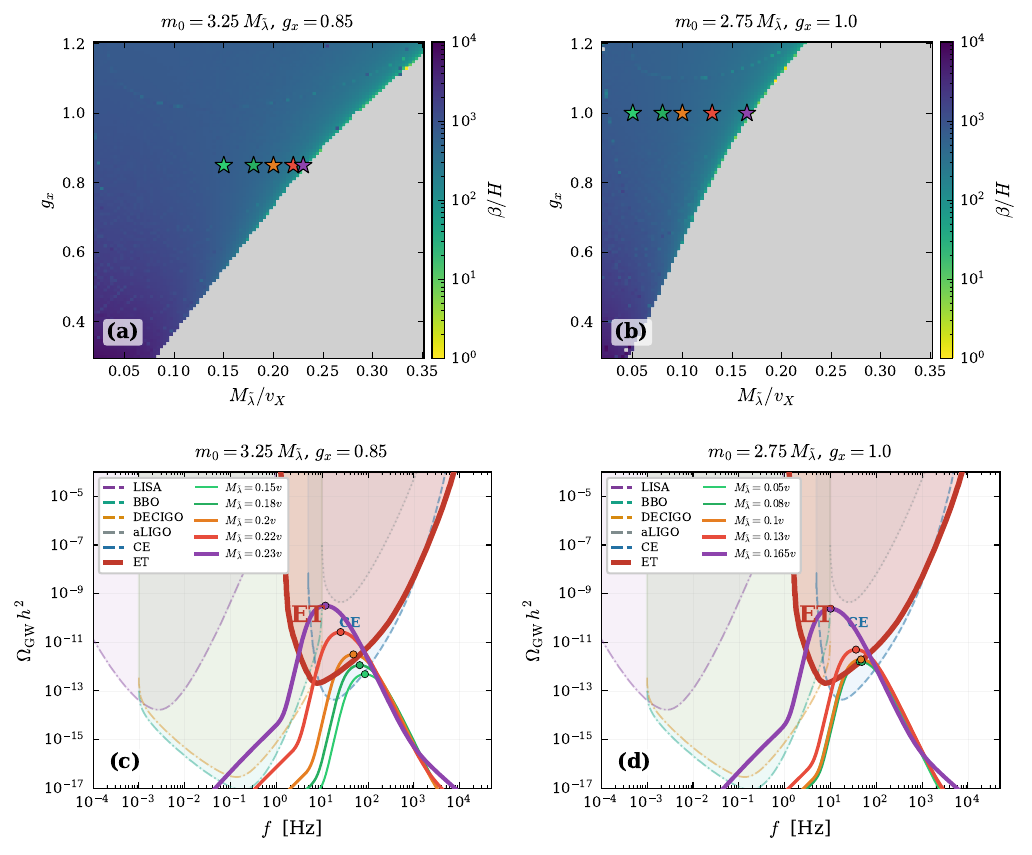}
  \caption{Parameter space and GW spectra for the SUSY flat-direction
  scenario ($\lambda_h = 0$, $v_X = 10^7\;\gev$, $T_h=T_v$).
  Left column: $m_0 = 3.25\,M_{\tilde\lambda}$, $g_x = 0.85$;
  right column: $m_0 = 2.75\,M_{\tilde\lambda}$, $g_x = 1.0$.
  (a,b): $\beta/H$ in the $g_x$--$M_{\tilde\lambda}/v_X$
  plane; gray: percolation fails;
  colored stars: five benchmarks shown below.
  (c,d): GW power spectra for the five benchmarks (solid),
  with projected sensitivities of LISA, BBO, DECIGO, aLIGO, ET,
  and CE.
  Decreasing $m_0/M_{\tilde\lambda}$ shrinks the viable region
  but deepens the supercooling near the percolation boundary.}
  \label{fig:Mhalf_scan}
  \label{fig:GW_spectra}
  \end{figure}

Throughout this section the hidden and visible sectors are taken at a common
temperature, $T_h=T_v$; the case $T_h\neq T_v$ is treated in the next section.
The first order phase transition proceeds via bubble
nucleation~\cite{Coleman:1977py,Callan:1977pt,Coleman:1980aw,Linde:1980tt,Linde:1981zj}, with the
percolation temperature $\Tp$ defined by $P_f(\Tp) = 0.71$
and the mean bubble separation $\Rstar$ computed from the
bounce action $S_3(T)$ obtained with
CosmoTransitions~\cite{Wainwright:2011kj} (see
also~\cite{Guada:2020xnz,Basler:2018cwe}).
In the analysis we use the mean bubble separation $\Rstar$
rather than the usual parametrization $\beta_*/H_*$, since
$\beta_*/H_*$ becomes inaccurate for strongly supercooled
transitions~\cite{Li:2025nja}; the precise definitions of
$\beta_*/H_*$ and $\Rstar$ and their relationship are given
in Appendix~\ref{sec:appendix_PT} (see the text after
Eq.~(\ref{eq:Rstar_app})). In the flat-direction ($\lambda_h = 0$), the parameter space
is controlled by $M_{\tilde\lambda}/v_X$, the ratio
$m_0/M_{\tilde\lambda}$, and $g_x$.
We present results for two representative configurations
(Fig.~\ref{fig:Mhalf_scan} and Table~\ref{tab:benchmarks}):
(i)~$m_0 = 3.25\,M_{\tilde\lambda}$, $g_x = 0.85$, for which
mild supercooling occurs at
$M_{\tilde\lambda}/v_X \lesssim 0.10$, strong (ET-detectable)
supercooling at
$0.10 \lesssim M_{\tilde\lambda}/v_X \lesssim 0.23$,
and percolation fails above $\sim 0.23\,v_X$;
(ii)~$m_0 = 2.75\,M_{\tilde\lambda}$, $g_x = 1.0$, which
gives a smaller viable region (percolation boundary at
$\sim 0.17\,v_X$) but deeper supercooling near the boundary.
Both configurations produce peak GW amplitudes
$\Omega_{\rm GW} h^2 \simeq 2$--$3\times 10^{-10}$
in the ET band ($f \sim 10$--$12\;$Hz), well above
the sensitivity floor of $\sim 10^{-13}$.
As anticipated in Section~\ref{sec:vcw}, increasing $M_{\tilde\lambda}$ deepens
the radiative barrier, lowering $\Tp$ and strengthening $\atot$, so the
supercooling grows with $M_{\tilde\lambda}$ until percolation fails. The
ET-detectable signal therefore spans a broad region of the
$g_x$--$M_{\tilde\lambda}/v_X$ plane (Fig.~\ref{fig:Mhalf_scan}a,b).

\begin{table}[tbp]
\centering
\small
\begin{tabular}{l|ccccccc}
\hline
Benchmark
  & $T_c/v_X$ & $T_p/v_X$ & $T_p/T_c$
  & $\atot$ & $\beta/H$ & $f_{\rm peak}$ [Hz] & $\Omega_{\rm GW} h^2$ \\
\hline
\multicolumn{8}{c}{\textit{$m_0 = 3.25\,M_{\tilde\lambda}$,
  $g_x = 0.85$ (Fig.~\ref{fig:Mhalf_scan}c)}} \\
\hline
$M_{\tilde\lambda} = 0.15\,v_X$
  & 0.140 & 0.069 & 0.492 & 0.097 & 693 & 85
  & $4.9\!\times\! 10^{-13}$ \\
$M_{\tilde\lambda} = 0.18\,v_X$
  & 0.147 & 0.068 & 0.461 & 0.124 & 538 & 66
  & $1.1\!\times\! 10^{-12}$ \\
$M_{\tilde\lambda} = 0.20\,v_X$
  & 0.150 & 0.063 & 0.419 & 0.178 & 411 & 48
  & $3.2\!\times\! 10^{-12}$ \\
$M_{\tilde\lambda} = 0.22\,v_X$
  & 0.152 & 0.052 & 0.339 & 0.403 & 239 & 25
  & $2.6\!\times\! 10^{-11}$ \\
$M_{\tilde\lambda} = 0.23\,v_X$
  & 0.153 & 0.038 & 0.251 & 1.303 & 112 & 12
  & $3.2\!\times\! 10^{-10}$ \\
$M_{\tilde\lambda} \geq 0.235\,v_X$ & \multicolumn{7}{c}{percolation fails} \\
\hline
\multicolumn{8}{c}{\textit{$m_0 = 2.75\,M_{\tilde\lambda}$,
  $g_x = 1.0$ (Fig.~\ref{fig:Mhalf_scan}d)}} \\
\hline
$M_{\tilde\lambda} = 0.05\,v_X$
  & 0.648 & 0.033 & 0.051 & 0.243 & 706 & 43
  & $1.6\!\times\! 10^{-12}$ \\
$M_{\tilde\lambda} = 0.08\,v_X$
  & 0.124 & 0.044 & 0.353 & 0.184 & 598 & 48
  & $1.6\!\times\! 10^{-12}$ \\
$M_{\tilde\lambda} = 0.10\,v_X$
  & 0.131 & 0.048 & 0.368 & 0.180 & 532 & 46
  & $2.0\!\times\! 10^{-12}$ \\
$M_{\tilde\lambda} = 0.13\,v_X$
  & 0.140 & 0.050 & 0.359 & 0.222 & 382 & 36
  & $5.0\!\times\! 10^{-12}$ \\
$M_{\tilde\lambda} = 0.165\,v_X$
  & 0.149 & 0.038 & 0.256 & 0.890 & 105 & 10
  & $2.4\!\times\! 10^{-10}$ \\
$M_{\tilde\lambda} \geq 0.17\,v_X$ & \multicolumn{7}{c}{percolation fails} \\
\hline
\end{tabular}
\caption{Phase transition observables and GW power spectrum
with the expected sensitivities of the proposed detectors for
the benchmarks of Fig.~\ref{fig:Mhalf_scan} at
$v_X = 10^7\;\gev$, $T_h=T_v$ ($\overline{\rm DR}$ scheme).
First block: $m_0 = 3.25\,M_{\tilde\lambda}$,
$g_x = 0.85$; second block: $m_0 = 2.75\,M_{\tilde\lambda}$,
$g_x = 1.0$.
All benchmarks lie along the flat direction ($\lambda_h = 0$).
The two ``percolation fails'' rows give the threshold on
$M_{\tilde\lambda}/v_X$ above which the transition does not complete
in each configuration.
Near the percolation boundary, $\atot$ approaches or exceeds
unity.}
\label{tab:benchmarks}
\end{table}

We present ten SUSY flat-direction benchmarks in two
configurations: five at $m_0 = 3.25\,M_{\tilde\lambda}$,
$g_x = 0.85$ ($M_{\tilde\lambda}/v_X = 0.15$--$0.23$),
and five at $m_0 = 2.75\,M_{\tilde\lambda}$, $g_x = 1.0$
($M_{\tilde\lambda}/v_X = 0.05$--$0.165$), all at
$v_X = 10^7\;\gev$ and $T_h=T_v$
(Table~\ref{tab:benchmarks}).
In the next section we relax this assumption and allow $T_h\neq T_v$.
The GW power spectra are shown in
Fig.~\ref{fig:GW_spectra}(c,d), together with the projected
power-law integrated sensitivity curves~\cite{Schmitz:2020syl} of
LISA~\cite{LISA:2017pwj},
BBO~\cite{Corbin:2005ny}, DECIGO~\cite{Seto:2001qf,Kawamura:2011zz},
Taiji~\cite{Ruan:2018tsw}, TianQin~\cite{TianQin:2015yph},
$\mu$Ares~\cite{Sesana:2019vho}, aLIGO,
 Einstein Telescope (ET)~\cite{Punturo:2010zz}, and
Cosmic Explorer (CE)~\cite{Reitze:2019iox}.

In both configurations, the flat-direction benchmarks near the
percolation boundary produce the largest GW amplitudes
(Table~\ref{tab:benchmarks}). The strongest benchmarks
in each configuration ($M_{\tilde\lambda} = 0.23\,v_X$ at
$m_0 = 3.25$ and $M_{\tilde\lambda} = 0.165\,v_X$ at $m_0 = 2.75$)
both reach $\atot \approx 0.9$--$1.3$, signalling the onset
of ultra-strong supercooling with peak amplitudes
$\Omega_{\rm GW} h^2 \sim 2$--$3\times 10^{-10}$ at
$f \sim 10$--$12\;$Hz.
Stronger supercooling shifts the peak frequency downward: from
$\sim 85\;$Hz at the weakest benchmarks to $\sim 10\;$Hz at the
strongest. The ET sensitivity
floor is $\Omega_{\rm GW} h^2 \sim 10^{-13}$ at $f \sim 3$--$300\;$Hz. All
benchmarks in both configurations exceed this threshold,
with the strongest points giving
$\Omega_{\rm GW} h^2 \sim 2$--$3 \times 10^{-10}$, nearly three orders of magnitude
above the sensitivity floor.

\section{Coupled Thermal Evolution and Reheating Dynamics}
\label{sec:reheating}
\label{sec:coupled_evolution}

The gravitational-wave signal depends on both the phase-transition dynamics and
the thermal history of the hidden sector.  The kinetic-mixing portal $\delta$
sets the rate of energy exchange with the visible bath and thereby determines
the temperature ratio $\xi=T_h/T_v$.  Since $\xi$ fixes the hidden-sector share
of the total radiation density, and hence
$\alpha_{\rm tot}\propto\xi^4$, it directly controls the observable amplitude
and also enters the redshift of the spectrum.  A self-consistent evolution of
$\xi$ is therefore required; imposing $T_h=T_v$ would erase a physical parameter
of the hidden-sector history.

We describe the thermal history with a coupled Boltzmann system that evolves the
temperature ratio, the relevant hidden-sector yields, and the nucleation
history simultaneously.  For a strongly supercooled transition the released
vacuum energy also reheats the converted phase.  We treat this reheating with a
two-temperature description: bubbles nucleate in a cold false-vacuum exterior,
while latent heat is deposited in the true-vacuum interior.  This separation is
the key ingredient that identifies which temperature controls tunneling and
which temperature controls the reheated energy budget.  Related treatments of reheating during
first-order phase transitions have been developed in
Ref.~\cite{Matuszak:2026xsz}.

\subsection{The extended Boltzmann equations with vacuum energy and nucleation moments}
\label{sec:boltzmann_review}
\label{sec:vacuum_source}

During the transition the hidden sector is not described by a single
temperature.  The unconverted false vacuum outside the bubbles remains
supercooled, with temperature $T_{h,f}$, and controls the tunneling rate through
$\Gamma\propto e^{-S_3/T_{h,f}}$.  The converted true-vacuum interior has
temperature $T_{h,t}$ and receives the latent heat
$\Delta V(T_{h,f})=V_{\rm eff}(\phi_f,T_{h,f})-V_{\rm eff}(\phi_t,T_{h,f})$.
The visible sector remains at temperature $T_v$ and couples to the hidden sector
only through the portal currents $j_h,j_v\propto\delta^2$.  We therefore evolve
two ratios,
\begin{equation}
  \xi_f \equiv \frac{T_{h,f}}{T_v}\,, \qquad
  \xi_t \equiv \frac{T_{h,t}}{T_v}\,,
  \label{eq:xi_defs}
\end{equation}
which reduce to the usual single ratio $\xi$ when reheating is neglected.

We first recall the source-free evolution.  In the absence of latent-heat
release the hidden sector is a single bath governed by the two-sector energy
equations~\cite{Li:2025nja,Feng:2024pab},
\begin{align}
  \dot\rho_v+3H(\rho_v+p_v)&=j_v\,, \label{eq:boltz_v}\\
  \dot\rho_h+3H(\rho_h+p_h)&=j_h\,, \label{eq:boltz_h}
\end{align}
coupled only through the portal currents.  Rewriting these equations in terms of
$\xi_f$ gives the standard temperature-ratio equation~\cite{Li:2025nja}
\begin{equation}
    \frac{d\xi_f}{dT_v} = \frac{1}{T_v\,d\rho_h/dT_h}
    \left[-\xi_f\,\frac{d\rho_h}{dT_h}
    + \frac{4H\sigma_h\rho_h - j_h}
      {4H\sigma\rho - 4H\sigma_h\rho_h - j_v}
    \,\frac{d\rho_v}{dT_v}\right],
    \label{eq:dxi_dTv}
\end{equation}
with the bare hidden-sector heat capacity $d\rho_h/dT_h$.  Here
$\sigma_h=\tfrac34(1+p_h/\rho_h)$, while $\rho$ and $\sigma$ denote the
corresponding total quantities.  This source-free trajectory defines the cold
exterior temperature $T_{h,f}$.

The true-vacuum interior is obtained by adding the latent-heat source.  As the
converted fraction $\mathcal{F}$ grows, the released energy is deposited into
the hidden bath behind the bubble walls,
\begin{align}
  \dot\rho_v+3H(\rho_v+p_v)&=j_v\,, \label{eq:boltz_vR}\\
  \dot\rho_h+3H(\rho_h+p_h)&=j_h+\frac{d\mathcal{F}}{dt}\,\Delta V\,,
  \label{eq:boltz_hR}
\end{align}
which defines the reheated interior temperature $T_{h,t}$ and the ratio
$\xi_t$.  The corresponding temperature-ratio equation is
\begin{equation}
    \frac{d\xi_t}{dT_v} = \frac{1}{T_v\,d\rho_h/dT_h}
    \left[-\xi_t\,\frac{d\rho_h}{dT_h}
    + \frac{4H\sigma_h\rho_h - j_h - \dot{\mathcal F}\,\Delta V}
      {4H\sigma\rho - 4H\sigma_h\rho_h - j_v}
    \,\frac{d\rho_v}{dT_v}\right],
    \label{eq:dxit_dTv}
\end{equation}
where the source appears in the hidden energy flux.  Using the visible-sector
clock $\dot T_v=-HT_v$ and
$d\mathcal F/dT_v=\xi_f\,d\mathcal F/dT_{h,f}$ gives
$\dot{\mathcal F}\,\Delta V=-H\,T_{h,f}\,(d\mathcal F/dT_{h,f})\Delta V>0$.

This form makes energy conservation explicit.  If expansion and portal leakage
are neglected over the short duration of the transition, integrating
Eq.~(\ref{eq:boltz_hR}) gives
$\rho_h(\Treh)=\rho_h(T_p)+\Delta V$, the usual instantaneous-reheating limit.
In the dynamical treatment the source acts only while $\dot{\mathcal F}\neq0$.
After completion the interior ratio $\xi_t$ evolves adiabatically with the
reheating it has acquired, whereas the exterior ratio $\xi_f$ remains the
temperature relevant for nucleation.  Thus
Eqs.~(\ref{eq:boltz_v})--(\ref{eq:boltz_h}) define the cold branch, and
Eqs.~(\ref{eq:boltz_vR})--(\ref{eq:boltz_hR}) define the reheated branch.

The expansion rate is sourced by the volume-averaged total energy density.  It
contains visible radiation, hidden radiation in the false and true phases
weighted by $1-\mathcal{F}$ and $\mathcal{F}$, and the vacuum energy remaining
in the unconverted phase:
\begin{equation}
  H^2(T_v)=\frac{1}{3\Mpl^2}\Big[\rho_v
  +(1-\mathcal{F})\,\rho_{h,f}
  +\mathcal{F}\,\rho_{h,t}
  +(1-\mathcal{F})\,\Delta V\Big],
  \label{eq:Hubble_3sector}
\end{equation}
where $\rho_v=\tfrac{\pi^2}{30}g_v T_v^4$,
$\rho_{h,f}=\tfrac{\pi^2}{30}g_{h,f}(\xi_f T_v)^4$, and
$\rho_{h,t}=\tfrac{\pi^2}{30}g_{h,t}(\xi_t T_v)^4$.  The tunneling rate is
evaluated on the cold branch,
\begin{equation}
    \Gamma(T_{h,f}) = T_{h,f}^{\,4}
    \left(\frac{S_3(T_{h,f})}{2\pi T_{h,f}}\right)^{\!3/2}
    \exp\!\left(-\frac{S_3(T_{h,f})}{T_{h,f}}\right),
    \label{eq:Gamma_rate}
\end{equation}
and determines the converted fraction,
\begin{equation}
    \mathcal{F}(T_{h,f}) = 1 - \exp\!\left[-\frac{4\pi}{3}\,v_w^3
    \int_{T_{h,f}}^{T_{h,c}}
    \frac{dT'\;\Gamma(T')}{T'^{\,4}\,H(T')}
    \left(\int_{T'}^{T_{h,c}}\frac{dT''}{H(T'')}\right)^{\!3}
    \right],
    \label{eq:f_integral}
\end{equation}
with $\mathcal{F}=1$ corresponding to completion.

\label{sec:feedback}
The two-temperature construction separates two physical roles that coincide only
when reheating is negligible.  The tunneling rate depends on the cold exterior
temperature $T_{h,f}$, whereas the reheated interior temperature $T_{h,t}$
enters the energy budget and redshift.  We therefore do not use $\xi_t$ in the
exponent of $\Gamma$.

Since the latent-heat source changes the temperature evolution during
percolation, the false-vacuum fraction cannot be pre-computed independently of
the thermal system.  We therefore replace the nested integral by a hierarchy of
nucleation moments $U_n$, evolved in the visible-sector temperature variable
(Appendix~\ref{sec:appendix_moment_derivation}),
\begin{equation}
    \frac{dU_0}{dT_v} = -\frac{\xi_f\,\Gamma(T_{h,f})}{T_{h,f}^{\,4}\,H}\,,
    \qquad
    \frac{dU_n}{dT_v} = -\frac{n\,\xi_f\,U_{n-1}}{H}\,,
    \quad n = 1,2,3,
    \label{eq:moment_hierarchy}
\end{equation}
with $U_n(T_{v,c}) = 0$ and $\Gamma$ evaluated at $T_h=T_{h,f}$.  The converted
fraction and its derivative then follow algebraically,
\begin{equation}
    \mathcal{F} = 1 - e^{-I}\,, \quad
    I \equiv \frac{4\pi}{3}\,v_w^3\,U_3\,, \qquad
    \frac{d\mathcal{F}}{dT_{h,f}} = -(1-\mathcal{F})\,\frac{4\pi\,v_w^3\,U_2}{H}\,,
    \label{eq:I_from_moments}
\end{equation}
so $d\mathcal{F}/dT_{h,f}$ is determined by the current state, through $U_2$, without
implicit dependence on $d\xi_f/dT_v$.

\label{sec:full_system}%
The full dynamical state therefore has eleven components,
\begin{equation*}
\mathbf{y} = (\xi_f,\,\xi_t,\, Y_{\tilde\chi},\, Y_{\gamma'},\,
Y_\chi,\, Y_q,\, Y_{\sigma},\, U_0,\, U_1,\, U_2,\, U_3)\,,
\end{equation*}
comprising the two temperature ratios, the five hidden-sector yields, and the
four nucleation moments.  We integrate this system over $T_v$ in three stages:
ordinary two-sector evolution before nucleation, the full coupled system during
the transition, and ordinary evolution again after completion.  The single
Hubble rate in Eq.~(\ref{eq:Hubble_3sector}) interpolates continuously between
the vacuum-dominated and radiation-dominated regimes.  The explicit component
equations, including the comoving-distance Jacobian
$dh/dT_v=-\xi_f/H$, are given in Appendix~\ref{sec:appendix_boltzmann}.

\subsection{Dependence of the GW power spectrum on the portal coupling $\delta$
and temperature ratio $\xi$}
\label{sec:delta_dependence}

\begin{table}[t]
\centering
\small
\begin{tabular}{r|cccccccc}
\hline
$\delta$ & $\xi_{f,p}$ & $\xi_{t,p}$ & $\alpha_{\rm DS}$ & $\atot$
  & $\beta/H$ & $f_{\rm peak}$ [Hz]
  & $\Omega_{\rm GW} h^2$ \\
\hline
\multicolumn{8}{c}{\textit{FD benchmark
  ($M_{\tilde\lambda} = 0.23\,v_X$, $\xi_0 = 0.3$)}} \\
\hline
ent.\ cons. & 0.380 & -- & 177.0 & 0.077 & 91 & 1.3 & $7.1\!\times\!10^{-13}$ \\
$10^{-6}$ & 0.383 & 0.941 & 176.5 & 0.079 & 64 & 1.2 & $7.3\!\times\!10^{-13}$ \\
$10^{-5}$ & 0.496 & 1.198 & 163.4 & 0.206 & 66 & 1.0 & $3.6\!\times\!10^{-12}$ \\
$10^{-4}$ & 1.012 & 2.395 & 143.2 & 3.034 & 70 & 0.9 & $7.1\!\times\!10^{-11}$ \\
\hline
\multicolumn{8}{c}{\textit{FD benchmark
  ($M_{\tilde\lambda} = 0.23\,v_X$, $\xi_0 = 1$)}} \\
\hline
ent.\ cons. & 1.267 & -- & 141.9 & 7.093 & 100 & 11.5 & $1.2\!\times\!10^{-10}$ \\
$10^{-6}$ & 1.176 & 2.739 & 141.8 & 5.357 & 72 & 9.4 & $1.2\!\times\!10^{-10}$ \\
$10^{-5}$ & 1.176 & 2.745 & 141.6 & 5.350 & 71 & 9.3 & $1.2\!\times\!10^{-10}$ \\
$10^{-4}$ & 1.015 & 2.404 & 143.1 & 3.074 & 70 & 0.9 & $7.2\!\times\!10^{-11}$ \\
\hline
\end{tabular}
\caption{Phase-transition observables and gravitational-wave power spectrum for
the near-boundary flat-direction benchmark $M_{\tilde\lambda} = 0.23\,v_X$
($v_X = 10^7\;\gev$, $g_x = 0.85$, $m_0 = 3.25\,M_{\tilde\lambda}$), computed
with the two-temperature solver at $\xi_0 = 0.3$ (upper block) and $\xi_0 = 1$
(lower block). $\xi_{f,p}$ and $\xi_{t,p}$ are the cold false-vacuum and reheated
true-vacuum ratios at percolation (the final reheating ratio is larger; see
Table~\ref{tab:Treh_compare}); the ``ent.\ cons.'' rows are the decoupled
($\delta\to0$) entropy-conservation baseline (no reheated interior, so
$\xi_{t,p}$ does not apply). The hot weak-portal cases ($\xi_0=1$,
$\delta\leq10^{-5}$) are sound-wave dominated and peak in the ET band near
$\sim\!10\;$Hz; the strongly supercooled cases that fall in the runaway/collision
regime (the cold $\xi_0=0.3$ block and the $\delta=10^{-4}$ rows) have their peak
set by the bubble-collision (envelope) source near $\sim\!1\;$Hz, and there the
quoted $\Omega_{\rm GW}h^2$ should be read as indicative.}
\label{tab:reheating_results}
\end{table}

\begin{figure}[t]
  \centering
  \includegraphics[width=\linewidth]{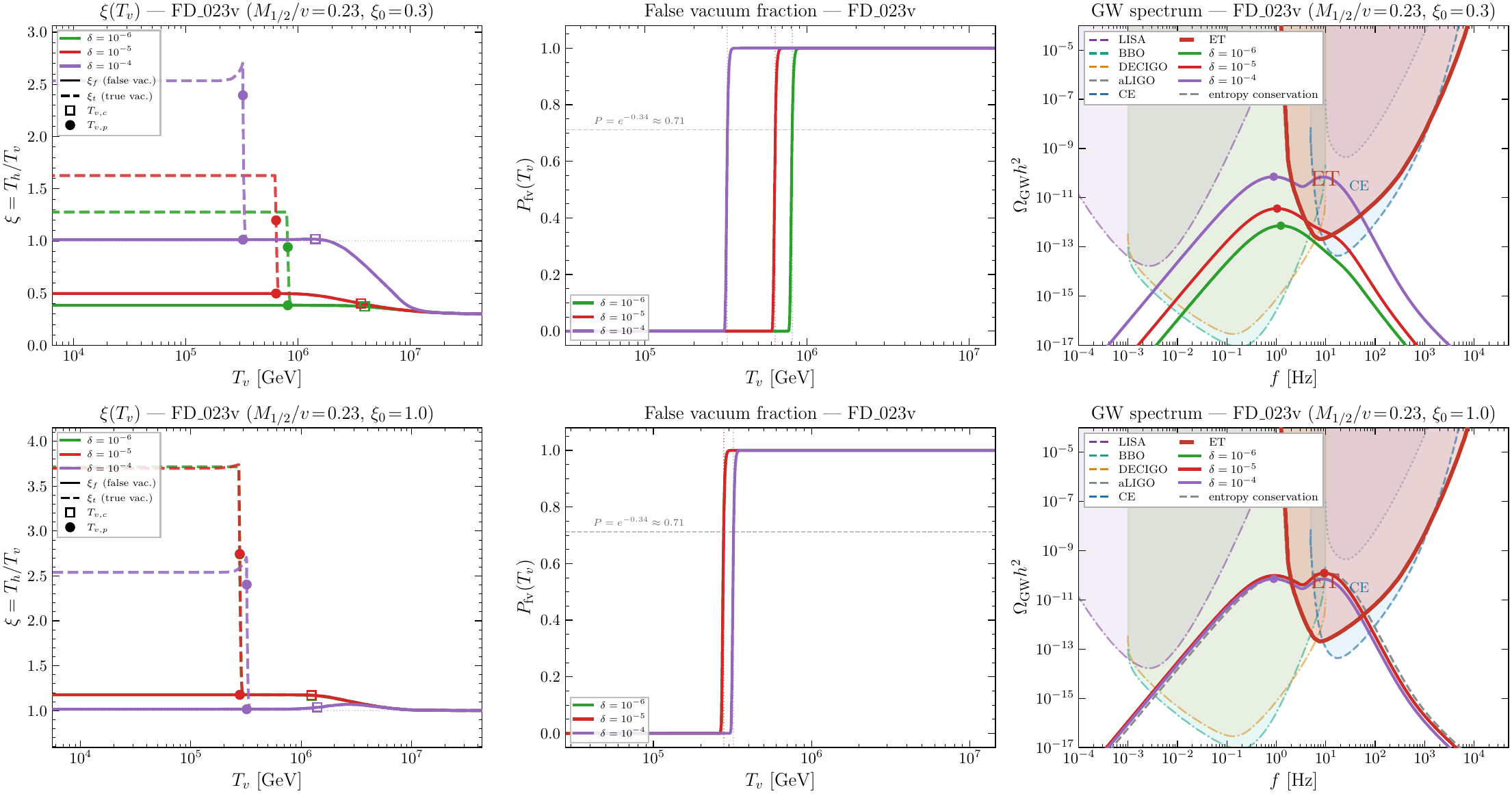}
  \caption{Temperature evolution, false-vacuum fraction, and GW spectrum for
  the near-boundary benchmark $M_{\tilde\lambda}/v_X=0.23$.  The columns
  correspond to $\delta=10^{-6},10^{-5},10^{-4}$, while the upper and lower rows
  show $\xi_0=0.3$ and $\xi_0=1$, respectively.  In the temperature panels, the
  solid curve is the cold false-vacuum ratio $\xi_f$ and the dashed curve is the
  reheated true-vacuum ratio $\xi_t$.  The grey dashed curves in the GW panels
  denote the entropy-conservation baseline.  The peak frequency is
  regime-dependent: the hot weak-portal cases ($\xi_0=1$, $\delta\leq10^{-5}$)
  are sound-wave dominated and peak in the ET band near $\sim\!10\;$Hz, while the
  strongly supercooled cases (the $\xi_0=0.3$ block and the $\delta=10^{-4}$
  rows) reach the runaway/collision regime and peak near $\sim\!1\;$Hz.  The
  amplitude is primarily set by the cold percolation ratio $\xi_{f,p}$ listed in
  Table~\ref{tab:reheating_results}.}
  \label{fig:GW_reheating}
  \end{figure}

The observable GW signal depends sensitively on the hidden-sector temperature
ratio at percolation.  In the present setup this ratio is determined by two
quantities: the portal coupling $\delta$, which controls energy exchange between
the visible and hidden baths, and the initial value
$\xi_0\equiv\xi(T\!\gg\!T_c)$, which encodes the earlier thermal history.  We
therefore apply the two-temperature evolution of Section~\ref{sec:full_system}
to the near-boundary flat-direction benchmark
$M_{\tilde\lambda}=0.23\,v_X$ with $v_X=10^7\;\gev$, scanning
$\delta=10^{-6},10^{-5},10^{-4}$ for $\xi_0=0.3$ and $\xi_0=1$.
The corresponding transition parameters and GW observables are listed in
Table~\ref{tab:reheating_results}.  Figure~\ref{fig:GW_reheating} displays the
temperature evolution, false-vacuum fraction, and redshifted GW spectrum for
the same cases.
  
The transition dynamics are largely fixed by the effective potential and by the
cold-exterior nucleation history.  Consequently the hidden-sector strength and
duration vary only mildly over the scan:
$\alpha_{\rm DS}\simeq140$--$180$ and $\beta/H\simeq65$--$100$.  This behavior
is visible in the middle column of Fig.~\ref{fig:GW_reheating}, where the
false-vacuum fraction evolves in nearly the same way when expressed in terms of
the cold nucleating temperature.  The main role of $\delta$ and $\xi_0$ is
instead to determine the cold percolation ratio
$\xi_{f,p}\equiv\xi_f(\Tp)$.  This ratio controls the fraction of the total
radiation density stored in the hidden sector and hence the observable strength,
\begin{equation}
  \alpha_{\rm tot} = \alpha_{\rm DS}\,
  \frac{\rho_{\rm rad}^h}{\rho_{\rm rad}^{\rm tot}} \propto \xi_{f,p}^4\,,
  \label{eq:atot_xi4}
\end{equation}
which sets the GW amplitude.

For the colder initial condition, $\xi_0=0.3$, increasing $\delta$ heats the
hidden sector toward the visible bath.  At $\delta=10^{-6}$ the portal is still
inefficient, giving $\xi_{f,p}=0.383$ and
$\alpha_{\rm tot}=0.079$; the peak amplitude is
$\Omega_{\rm GW}h^2=7.3\times10^{-13}$.  At $\delta=10^{-5}$ the portal raises
$\xi_{f,p}$ to $0.496$, increasing $\alpha_{\rm tot}$ to $0.206$ and the peak
to $3.6\times10^{-12}$.  For $\delta=10^{-4}$ the sectors are close to thermal
contact at percolation, $\xi_{f,p}=1.012$, and the $\xi_{f,p}^4$ suppression is
removed; the peak then reaches $7.1\times10^{-11}$.

The hotter initial condition, $\xi_0=1$, gives an observable signal even for a
weak portal.  In this case $\xi_{f,p}\simeq1.18$ for
$\delta=10^{-6}$ and $10^{-5}$, yielding
$\alpha_{\rm tot}\simeq5.35$ and
$\Omega_{\rm GW}h^2\simeq1.2\times10^{-10}$.  At
$\delta=10^{-4}$ both initial conditions converge, as the portal drives
$\xi_{f,p}$ to unity and the two spectra coincide within the accuracy shown in
Table~\ref{tab:reheating_results}.  Thus $\xi_0$ is not a removable convention:
unless the portal equilibrates the two sectors before percolation, the initial
hidden temperature remains imprinted on the GW amplitude.

The peak frequency is regime-dependent.  The hot weak-portal cases ($\xi_0=1$,
$\delta\leq10^{-5}$) are sound-wave dominated and peak in the Einstein Telescope
band at $f_{\rm peak}\simeq9\;$Hz, close to the entropy-conservation baseline
($11.5\;$Hz)---a direct illustration that reheating leaves the spectrum nearly
unchanged.  The remaining cases---the entire $\xi_0=0.3$ block and the
$\delta=10^{-4}$ rows---instead fall in the runaway/collision regime, where the
spectrum is dominated by the bubble-collision (envelope) source and peaks near
$f_{\rm peak}\simeq1\;$Hz, at the low-frequency edge of the band.  All cases
remain at or above the ET floor, $\Omega_{\rm GW}h^2\gtrsim10^{-13}$.

\subsection{Reheating effects on the GW spectrum}
\label{sec:reheating_gw}

\begin{table}[t]
\centering
\small
\begin{tabular}{r|cccc}
\hline
$\delta$ & $\xi_{f,p}$ & $\xi_{t,p}$
  & $\xi_{t,{\rm completion}}$ & $\xi_{\rm reh}^{\rm inst.}$ \\
\hline
\multicolumn{5}{c}{\textit{$M_{\tilde\lambda} = 0.23\,v_X$, $\xi_0 = 0.3$}} \\
\hline
$10^{-6}$ & 0.383 & 0.941 & 1.277 & 1.287 \\
$10^{-5}$ & 0.496 & 1.198 & 1.626 & 1.637 \\
$10^{-4}$ & 1.012 & 2.395 & 2.534 & 3.231 \\
\hline
\multicolumn{5}{c}{\textit{$M_{\tilde\lambda} = 0.23\,v_X$, $\xi_0 = 1$}} \\
\hline
$10^{-6}$ & 1.176 & 2.739 & 3.715 & 3.747 \\
$10^{-5}$ & 1.176 & 2.745 & 3.698 & 3.746 \\
$10^{-4}$ & 1.015 & 2.404 & 2.539 & 3.242 \\
\hline
\end{tabular}
\caption{Cold- and true-vacuum hidden-sector temperature ratios for the
near-boundary benchmark $M_{\tilde\lambda}=0.23\,v_X$, for the portal strengths
used in Table~\ref{tab:reheating_results}.  The cold exterior ratio
$\xi_{f,p}$ controls bubble nucleation, while $\xi_{t,p}$ is the reheated
true-vacuum ratio at percolation.  The column $\xi_{t,{\rm completion}}$ gives
the final true-vacuum ratio after completion, and
$\xi_{\rm reh}^{\rm inst.}$ is the instantaneous estimate obtained from
$\rho_h(T_{\rm reh})=\rho_h(T_p)+\Delta V$.  The percent-level agreement at
$\delta\leq10^{-5}$ shows that the weak-portal transition is effectively
isolated, whereas the $\delta=10^{-4}$ rows show the reduction from energy
transfer to the visible bath.}
\label{tab:Treh_compare}
\end{table}

\begin{figure}[t]
  \centering
  \includegraphics[width=0.48\linewidth]{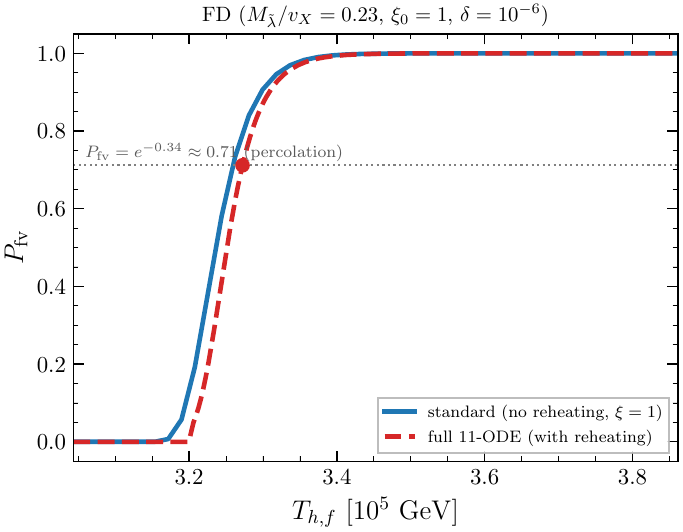}
  \caption{False-vacuum fraction $P_{\rm fv}$ as a function of the cold
  exterior temperature $T_{h,f}$ for the near-boundary benchmark
  $M_{\tilde\lambda}/v_X=0.23$ at $\xi_0=1$ and $\delta=10^{-6}$.  The solid
  curve is the standard calculation without reheating, while the dashed curve
  is the two-temperature result including latent-heat release.  Since
  $\Gamma\propto\exp[-S_3(T_{h,f})/T_{h,f}]$ is controlled by the cold exterior,
  both treatments percolate at essentially the same temperature,
  $T_{h,f}\simeq3.3\times10^5\;\gev$ ($P_{\rm fv}=e^{-0.34}$, marked by the
  circle).  Reheating therefore changes the interior thermal history but leaves
  the nucleation history nearly unchanged.}
  \label{fig:Pfv_reheating}
  \end{figure}

We now quantify how the latent-heat release affects the final spectrum.  In the
two-temperature description the nucleation rate is controlled by the cold
false-vacuum exterior, with ratio $\xi_f$, while the released vacuum energy
heats the true-vacuum interior, with ratio $\xi_t$.  Table~\ref{tab:Treh_compare}
shows that this reheating is sizable: at percolation
$\xi_{t,p}/\xi_{f,p}\simeq2.3$--$2.5$, and after completion
$\xi_{t,{\rm completion}}/\xi_{f,p}\simeq2.5$--$3.3$.  For
$\delta=10^{-6}$ and $10^{-5}$ the completion value agrees with the
instantaneous estimate
$\rho_h(T_{\rm reh})=\rho_h(T_p)+\Delta V$ at the percent level, as expected for
a rapidly completing transition with negligible portal leakage.  For
$\delta=10^{-4}$ the instantaneous estimate is too high by about $22\%$, because
part of the released energy is transferred to the much larger visible bath.

The reheated temperature, however, is not the nucleation temperature.  The
exponential factor in the tunneling rate,
$\Gamma\propto\exp[-S_3(T_{h,f})/T_{h,f}]$, is evaluated in the unconverted
exterior.  This is seen directly in Fig.~\ref{fig:Pfv_reheating}: plotted
against $T_{h,f}$, the false-vacuum fractions with and without reheating nearly
coincide, and percolation occurs at the same
$T_{h,f}\simeq3.3\times10^5\;\gev$.  Thus the reheating changes the thermal
state of the converted phase, but leaves the percolation point and $\beta/H$
essentially fixed.

The remaining effect on the observed GW spectrum enters through the redshift
factors of Appendix~\ref{sec:appendix_gw_spectrum}, $\mathcal{R}_\Omega\propto
g_{\rm eff}h_{\rm eff}^{-4/3}$ and $\mathcal{R}_f\propto h_{\rm eff}^{-1/3}$, with
$g_{\rm eff}=g_{\rm eff}^v+g_{\rm eff}^h\xi_t^4$ and
$h_{\rm eff}=h_{\rm eff}^v+h_{\rm eff}^h\xi_t^3$.  The reheated interior
contributes to these factors through the $\xi_t^4$ weighting, but its net effect
on the spectrum is moderate: comparing the full two-temperature result with the
decoupled entropy-conservation baseline (Table~\ref{tab:reheating_results}), the
peak amplitude is essentially unchanged and the peak frequency shifts by
$\lesssim20\%$ (e.g.\ $11.5\to9.4\;$Hz for the $\xi_0=1$, $\delta=10^{-6}$ case).

Consequently the large separation between $\xi_f$ and $\xi_t$ is physically
important for the thermal history, but it is not a large uncertainty in the GW
prediction.  The robust prediction is obtained from the cold-exterior
nucleation history, with the reheated interior entering only through the energy
budget and the redshift factors above.

\section{Dark Matter and Cosmological Constraints}
\label{sec:cosmo_constraints}
\label{sec:dq_in_PT}
\label{sec:dq_spectrum}
\label{sec:heavy_relics}

\begin{figure}[t]
\centering
\includegraphics[width=\linewidth]{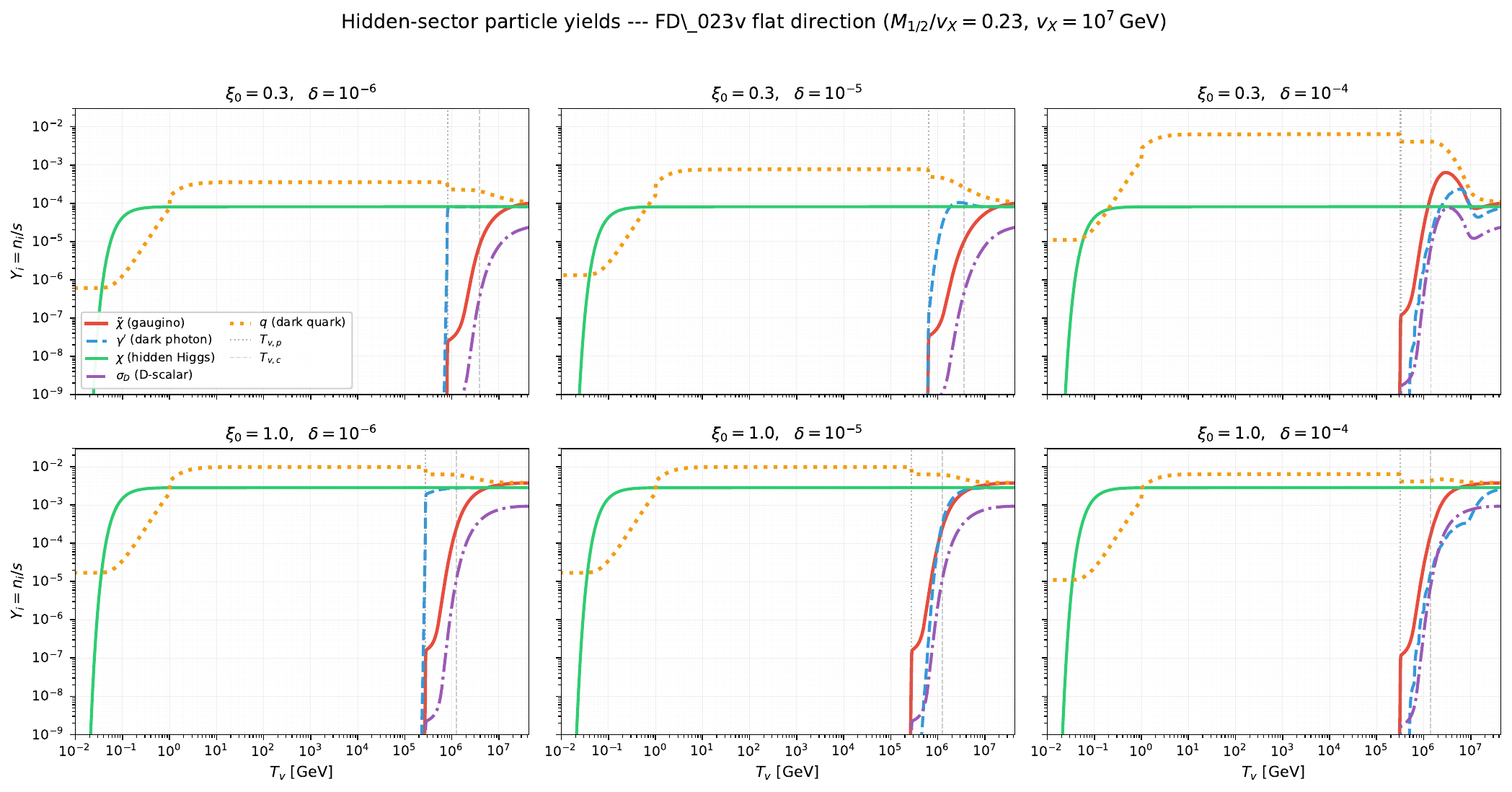}
\caption{Hidden-sector yields $Y_i=n_i/s$ as functions of the visible
temperature for the near-boundary benchmark
$M_{\tilde\lambda}/v_X=0.23$.  The columns correspond to
$\delta=10^{-6},10^{-5},10^{-4}$, and the upper and lower rows show
$\xi_0=0.3$ and $\xi_0=1$, respectively.  The gauge-sector states decay inside
the bubbles, $\chi$ decays before BBN, and the dark quark $q$ freezes out at the
phase transition and remains as the dark matter relic.}
\label{fig:yields}
\end{figure}

The lightest $U(1)_X$-charged state is the Dirac dark quark $q$, with
$g_q=4$.  Gauge invariance makes it stable, and it therefore provides the dark
matter candidate in this model.  Throughout the phase transition it is
ultrarelativistic, $m_q/T_h\lesssim10^{-8}$, so it contributes
$g_{*,q}^h=(7/8)\times4=3.5$\label{eq:gstar_dq} to the hidden relativistic
degrees of freedom, while its contribution to the effective potential is
negligible, $(y_q/g_x)^4\sim10^{-38}$.  The heavier gauge-sector states
($\gamma'$, $\sigma$, $\tilde\chi_\pm$) decay inside the broken-phase bubbles
with lifetimes $\tau\sim10^{-25}$--$10^{-30}\;$s, much shorter than a Hubble
time.  The remaining hidden Higgs state $\chi$ decays through
gravity-mediated interactions before BBN
(Appendix~\ref{sec:appendix_chi_decay}).  The corresponding yield evolution is
shown in Fig.~\ref{fig:yields}.

\subsection{Dark matter relic density}
\label{sec:relic_density}

\begin{table}[t]
  \centering
  \small
  \begin{tabular}{r|cccccc}
  \hline
  $\delta$ & $\xi_{f,p}$ & $Y_q^{\rm (dil)}$
    & $m_q$ [keV] & Ly-$\alpha$ margin
    & $\xi_{\rm BBN}$ & $\Neff$ \\
  \hline
  \multicolumn{7}{c}{\textit{FD benchmark
    ($M_{\tilde\lambda} = 0.23\,v_X$, $\xi_0 = 0.3$,
    $D = 584$, $\Treh = 98\;$MeV)}} \\
  \hline
  $10^{-6}$ & 0.383 & $5.8\!\times\!10^{-7}$
    & 752 & $1027\times$
    & $0.019$ & $2.7\!\times\!10^{-7}$ \\
  $10^{-5}$ & 0.496 & $1.2\!\times\!10^{-6}$
    & 356 & $375\times$
    & $0.025$ & $7.7\!\times\!10^{-7}$ \\
  $10^{-4}$ & 1.012 & $1.0\!\times\!10^{-5}$
    & 43 & $22\times$
    & $0.051$ & $1.3\!\times\!10^{-5}$ \\
  \hline
  \multicolumn{7}{c}{\textit{FD benchmark
    ($M_{\tilde\lambda} = 0.23\,v_X$, $\xi_0 = 1$,
    $D = 584$, $\Treh = 98\;$MeV)}} \\
  \hline
  $10^{-6}$ & 1.176 & $1.6\!\times\!10^{-5}$
    & 27 & $12\times$
    & $0.059$ & $2.4\!\times\!10^{-5}$ \\
  $10^{-5}$ & 1.176 & $1.6\!\times\!10^{-5}$
    & 27 & $12\times$
    & $0.059$ & $2.4\!\times\!10^{-5}$ \\
  $10^{-4}$ & 1.015 & $1.0\!\times\!10^{-5}$
    & 42 & $22\times$
    & $0.051$ & $1.4\!\times\!10^{-5}$ \\
  \hline
  \multicolumn{7}{l}{\textit{Bounds:}
    $\Omega_q h^2 = 0.12$ (fixes $m_q$); \;
    Ly-$\alpha$ margin ${>}\,1$~\cite{Viel:2013fqw}; \;
    $\Neff < 0.4$~\cite{Planck:2018vyg}} \\
  \hline
  \end{tabular}
  \caption{Dark quark relic abundance and cosmological constraints for the
  near-boundary benchmark $M_{\tilde\lambda}/v_X=0.23$.  The diluted yield
  $Y_q^{(\rm dil)}$ includes gauge-sector entropy redistribution
  ($\times1.57$) and the entropy release from $\chi$ decay
  ($\div D\simeq584$).  The mass $m_q$ is fixed by
  $\Omega_q h^2=0.12$.  The Ly-$\alpha$ margin is
  $m_q/[5.3\;\text{keV}\,(T_q/T_\nu)_0]$~\cite{Viel:2013fqw}; values larger
  than unity satisfy the bound.  All listed points also satisfy
  $\Neff\ll0.4$~\cite{Planck:2018vyg}.}
  \label{tab:cosmo_constraints}
  \end{table}
  
The relic abundance is obtained from the standard freeze-out
calculation~\cite{Scherrer:1985zt,Griest:1990kh,Gondolo:1990dk}; for reviews of
particle dark matter and its constraints see Ref.~\cite{Bertone:2004pz}.  Hidden
$U(1)_X$ sectors of this type have also been studied in the contexts of
self-interacting and asymmetric dark matter and matter--antimatter
asymmetry~\cite{Kaplan:2009ag,Zurek:2013wia,Shelton:2010ta,Haba:2010bm,Petraki:2013wwa,Fukugita:1986hr,Li:2023nez,Feng:2025wvc}.
The dark quark yield obeys Eq.~(\ref{eq:dYq_dTv}), with annihilation channels
$q\bar q\to\gamma'\gamma'$, $q\bar q\to\tilde\lambda\tilde\lambda$, and
$q\bar q\to\tilde H\tilde{\bar H}$; the corresponding cross sections are given
in Appendix~\ref{sec:appendix_dq_xsec}.  At $T_h=\Tp$ the total interaction rate
satisfies $\Gamma_{\rm tot}/H\simeq3\times10^8$
[Eq.~(\ref{eq:Gamma_over_H})], so $Y_q$ tracks its equilibrium value throughout
the symmetric phase.

Freeze-out occurs abruptly at the phase transition.  Once the gauge-sector
states acquire masses of order $g_xv_X\gg T_p$, all three annihilation channels
shut off.  The extended Boltzmann solver gives, for the
$M_{\tilde\lambda}=0.23\,v_X$, $\xi_0=0.3$, $\delta=10^{-6}$ benchmark,
$Y_q^{(0)}\simeq2.3\times10^{-4}$ at $\xi_{\rm fo}\simeq0.38$.  Subsequent
decays of $\gamma'$, $\sigma$, and $\tilde\chi_\pm$ redistribute the hidden
entropy into the dark quark bath, enhancing the yield by
$h_*^h/g_{*,q}\simeq5.5/3.5\simeq1.57$.

The final abundance is then diluted by the late decay of the hidden Higgs
$\chi$.  On the flat direction $\chi$ has no fast hidden decay channel and
decays through gravity-mediated interactions, with
$m_\chi=g_x^2m_0/(4\pi)\simeq4\times10^5\;\gev$,
$Y_\chi\sim10^{-4}$, and $\tau_\chi\sim10^{-5}\;$s
($T_{v,\rm dec}\sim98\;$MeV).  Before decaying, $\chi$ dominates the energy
density at $T_v\sim55\;\gev$, producing an early matter-dominated era.  The
entropy dilution factor $D\equiv s_{\rm after}/s_{\rm before}$ follows from
converting the rest-mass energy
$\rho_\chi=m_\chi n_\chi=m_\chi Y_\chi s_{\rm before}$ into radiation at
$\Treh$,
$s_{\rm after}=(4/3)\rho_\chi/\Treh$~\cite{Scherrer:1985zt,Kolb:1990vq,Moroi:1999zb,Nemevsek:2023yjl,Giudice:2000ex},
which gives

\begin{align}
    D = \frac{4}{3}\,\frac{m_\chi\,Y_\chi}{\Treh}\,.
    \label{eq:entropy_dilution}
\end{align}
For the benchmark above the numerical solution gives $D\simeq584$ and
$\Treh\simeq98\;$MeV.  The resulting diluted yield is
$Y_q^{(\rm dil)}\simeq5.8\times10^{-7}$, and the observed abundance
$\Omega_qh^2=0.12$ fixes $m_q\simeq752\;$keV.  The dilution also lowers the dark
quark momentum by $D^{1/3}\simeq8.4$, so the Lyman-$\alpha$
free-streaming bound~\cite{Viel:2013fqw,Irsic:2017ixq} is easily satisfied
throughout Table~\ref{tab:cosmo_constraints}.  The low-$\xi_0$ cases have a
smaller total transition strength, but the corresponding GW signal remains in
the ET sensitivity range, as shown in Table~\ref{tab:reheating_results}.

\subsection{BBN constraint on extra radiation}
\label{sec:BBN}

At $T_v\sim1\;\text{MeV}$ any remaining hidden radiation contributes to
$\Neff$.  Normalizing to one neutrino species,
$\rho_{\nu,1}=(7/8)\,2\,(\pi^2/30)T_\nu^4$, the hidden contribution
$\rho_h=(\pi^2/30)g_*^hT_h^4$ gives
$\Neff=(4/7)g_*^h(T_h/T_\nu)^4$.  Since $T_\nu\simeq T_v$ at BBN,
\begin{align}
    \Neff = \frac{4}{7}\,g_*^h(T_{h,\rm BBN})\,\xi_{\rm BBN}^4\,.
    \label{eq:Neff}
\end{align}
Only the dark quark remains relativistic at this epoch, so $g_*^h=3.5$; the
other hidden states have masses above $10^5\;\gev$.  The hidden-to-visible
temperature ratio is strongly reduced by the visible-sector entropy release
from the MSSM to the SM, by the redistribution of gauge-sector entropy into the
dark quark bath, and by the late $\chi$ entropy dilution
($D^{-1/3}\simeq0.12$).  For example, the Boltzmann solution gives
$\xi_{\rm BBN}=0.019$ for $\xi_0=0.3$ and $\delta=10^{-6}$, which corresponds to
$\Neff\simeq(4/7)\,3.5\,(0.019)^4\simeq3\times10^{-7}$.
This is far below the observational bound
$\Neff\lesssim0.4$~\cite{Planck:2018vyg,Fields:2019pfx,Cyburt:2015mya,Brust:2013ova,Aboubrahim:2022gjb}.
The largest value in Table~\ref{tab:cosmo_constraints}, obtained in the hot
$\xi_0=1$ block, is only $\Neff\simeq2.4\times10^{-5}$.  Hence all benchmarks
satisfy the BBN constraint by several orders of magnitude.

\section{Conclusion}
\label{sec:conclusion}

We have studied supercooled first-order phase transitions in a supersymmetric
hidden sector with a spontaneously broken $U(1)_X$ gauge symmetry.  Along the
D-flat direction the tree-level quartic vanishes, and the transition barrier is
generated by the Coleman--Weinberg potential induced by soft SUSY-breaking mass
splittings.  In the $\overline{\rm DR}$ scheme the supersymmetric contribution
to the effective potential cancels in the soft-unbroken limit, so the barrier is
controlled by the gaugino soft mass $M_{\tilde\lambda}$, while the soft scalar
mass $m_0$ stabilizes the broken vacuum.  A scan over the flat-direction
parameter space gives a broad viable window,
$M_{\tilde\lambda}/v_X\simeq0.05$--$0.23$, with no tuning of a tree-level
quartic.  Near the percolation boundary the signal reaches
$\Omega_{\rm GW}h^2\sim3\times10^{-10}$, well above the Einstein Telescope
sensitivity floor and in the frequency range relevant for ET and Cosmic
Explorer.

The observable amplitude is strongly affected by the hidden-sector thermal
history.  The portal coupling $\delta$ controls energy exchange between the
visible and hidden baths, while the initial ratio $\xi_0$ fixes the prior hidden
temperature.  These quantities determine the cold percolation ratio
$\xi_{f,p}$ and hence the hidden-sector share of the total radiation density.
For the near-boundary benchmark, increasing $\delta$ from $10^{-6}$ to
$10^{-4}$ raises the signal from the ET floor to
$\Omega_{\rm GW}h^2\simeq7\times10^{-11}$ for $\xi_0=0.3$, while a hotter initial
hidden sector gives a large signal already at weak portal coupling.  Thus
$\delta$ and $\xi_0$ are genuine cosmological inputs to the prediction.

We followed the transition with an 11-variable Boltzmann system that evolves the
cold exterior ratio $\xi_f$, the reheated interior ratio $\xi_t$, five
hidden-sector yields, and four nucleation moments.  The tunneling rate is
evaluated on the cold branch, $\Gamma=\Gamma(T_{h,f})$, whereas the reheated
true-vacuum branch enters through the volume-averaged energy density,
completion temperature, and redshift factors.  The latent heat can produce a
large split between $\xi_f$ and $\xi_t$, but for the benchmarks studied here it
leaves the cold-branch percolation history nearly unchanged and modifies the
final spectrum mainly through modest redshift corrections.

The same hidden sector can also reproduce the observed dark matter abundance.
The dark quark freezes out while relativistic and is subsequently diluted by the
entropy released in the late decay of the hidden Higgs, with
$D\simeq584$.  Across the benchmark set, the relic-density condition
$\Omega_{\rm CDM}h^2=0.12$ is obtained for
$m_q\simeq30$--$800\;$keV, with Lyman-$\alpha$ bounds comfortably satisfied.
The remaining dark radiation is far below current BBN limits,
$\Neff\lesssim{\rm few}\times10^{-5}$.  These results show that soft
SUSY-breaking parameters, the portal coupling, and the hidden-sector thermal
history together determine the GW signal and its cosmological consistency.

\section*{Acknowledgments}
The research was supported in part by the NSF Grant PHY-2209903.

\appendix
\section{The Coleman--Weinberg One-Loop Potential along the Flat Direction}
\label{sec:appendix_potential}

This appendix derives the one-loop Coleman--Weinberg (CW) potential of the
massive $\mathcal{N}=1$ vector supermultiplet along the $D$-flat direction in a
general $R_{\xi_{\rm gauge}}$ gauge, establishes the boson--fermion degree-of-freedom
counting that underlies the cancellation in the supersymmetric limit, and gives the
gaugino-induced logarithmic barrier used in Sections~\ref{sec:model}
and~\ref{sec:supercooling}. To match the conventions of
Section~\ref{sec:model}, we use the canonically normalized radial field
$\phi_c$, defined by $\Phi=\bar\Phi=\phi_c/2$. With this normalization the
gauge-boson mass is $m_A^2=g_x^2\phi_c^2$, as in
Table~\ref{tab:masses}, and the off-diagonal entry of the fermion mass matrix is
$m_A=g_x\phi_c$, as in
Eq.~(\ref{eq:ferm_mass_matrix}). All
remaining symbols---the gauge coupling $g_x$, the gaugino soft mass
$M_{\tilde\lambda}$, and the renormalization scale $\mu_R$---are exactly those of
Section~\ref{sec:model}.
\nw{We keep the singlet coupling $\lambda_S$ general here---the flat-direction
limit $\lambda_S=0$ is taken only in Section~\ref{sec:supercooling}---and use
the notation of Section~\ref{sec:model}: the quartic
$\lambda_h=\lambda_S^2/4$, the soft scalar mass $m_0^2$, and the curvature
parameter $m^2\equiv\lambda_S^2 v_X^2-m_0^2$ of Eq.~(\ref{eq:Vtree_flat}).} The fermion notation is also that
of Section~\ref{sec:model}: $\tilde H_\pm=(\tilde\Phi\pm\tilde{\bar\Phi})/\sqrt2$
are the gauge-basis Higgsino combinations (with $\tilde H_-$ mixing with the
gaugino and $\tilde H_+$ massless), and $\tilde\chi_\pm$ are the two Majorana
mass eigenstates with masses $m_\pm$.

We focus on the kinetic energy part of the lagrangian which involves  the gauge fields
\begin{equation}
  \lag_{\rm kin}= |D_\mu \Phi|^2+|D_\mu \bar\Phi|^2
  -\frac14 F_{\mu\nu}F^{\mu\nu},
\end{equation}
with
\begin{equation}
  D_\mu \Phi=(\partial_\mu-i g_x A_\mu)\Phi,
  \qquad
  D_\mu \bar\Phi=(\partial_\mu+i g_x A_\mu)\bar\Phi.
\end{equation}
Along the $D$-flat direction take the real background
\begin{equation}
  \Phi=\bar\Phi=\frac{\phi_c}{2}.
\end{equation}
It is useful to parametrize fluctuations as
\begin{align}
  \Phi&=\frac{1}{\sqrt2}
  \left[
  \frac{\phi_c}{\sqrt2}+\frac{h+\sigma}{\sqrt2}
  +i\frac{G_h^0+G}{\sqrt2}
  \right],\\
  \bar\Phi&=\frac{1}{\sqrt2}
  \left[
  \frac{\phi_c}{\sqrt2}+\frac{h-\sigma}{\sqrt2}
  +i\frac{G_h^0-G}{\sqrt2}
  \right].
\end{align}
The pseudoscalar combination which shifts under the broken $U(1)_X$ is the
antisymmetric imaginary part $G$, while $G_h^0$ is the orthogonal physical
pseudoscalar.  With these conventions the field-dependent
gauge-boson mass is
\begin{equation}
  {\mx^2(\phi_c)=g_x^2\phi_c^2.}
\end{equation}
This is the physical mass squared of the massive vector field $A_\mu$ in the broken
background.\\

Expanding the scalar kinetic terms to quadratic order in fluctuations gives, among
other terms,
\begin{equation}
  |D_\mu \Phi|^2+|D_\mu \bar\Phi|^2
  \supset
  \frac12 \mx^2 A_\mu A^\mu
  +\frac12 (\partial_\mu G)^2
  +\mx A_\mu\partial^\mu G.
\end{equation}
Equivalently, after integrating by parts,
\begin{equation}
  \mx A_\mu\partial^\mu G
  =-\mx G\,\partial_\mu A^\mu,
\end{equation}
up to a total derivative.  This is the familiar gauge-Goldstone mixing term.  It is
not convenient for computing propagators or the one-loop determinant, because the
quadratic operator is not diagonal in $A_\mu$ and $G$.
The $R_{\xi_{\rm gauge}}$ gauge is chosen precisely to cancel this mixing.  Use the gauge-fixing
functional
  $F[A,G]=\partial_\mu A^\mu-\xi_{\rm gauge} \mx G$, we  add the gauge fixing term
\begin{equation}
  {
  \lag_{\rm GF}=-\frac{1}{2\xi_{\rm gauge}}\left(\partial_\mu A^\mu-\xi_{\rm gauge} \mx G\right)^2 .
  }
\end{equation}
Expanding this term gives
\begin{align}
  \lag_{\rm GF}
  &=-\frac{1}{2\xi_{\rm gauge}}(\partial_\mu A^\mu)^2
  +\mx G\,\partial_\mu A^\mu
  -\frac12\xi_{\rm gauge} \mx^2 G^2 .
\end{align}
The middle term cancels the integrated-by-parts kinetic mixing,
\begin{equation}
  -\mx G\partial_\mu A^\mu
  +\mx G\partial_\mu A^\mu=0.
\end{equation}
The last term is the origin of the gauge-dependent Goldstone mass contribution,
\begin{equation}
  {\Delta m_G^2=\xi_{\rm gauge} \mx^2(\phi_c).}
\end{equation}
Using the above  the quadratic action is diagonalized and has the form
\begin{equation}
  \lag_A^{(2)}=\frac12 A_\mu
  \left[
  g^{\mu\nu}\left(\Box+\mx^2\right)
  -\left(1-\frac{1}{\xi_{\rm gauge}}\right)\partial^\mu\partial^\nu
  \right] A_\nu .
\end{equation}
In momentum space this operator has three physical massive vector eigenmodes with
mass $\mx^2$, and one scalar-like unphysical eigenmode with mass $\xi_{\rm gauge}\mx^2$.  This implies that one may write the gauge contribution to the one-loop effective potential as
\begin{equation}
  V_A^{(1)}(\phi_c)=\frac{1}{64\pi^2}
  \left\{
  3\mx^4\left[\ln\frac{\mx^2}{\mu_R^2}-\frac32\right]
  +\xi_{\rm gauge}^2\mx^4\left[\ln\frac{\xi_{\rm gauge}\mx^2}{\mu_R^2}-\frac32\right]
  \right\}.
\end{equation}
\nw{We work throughout in the $\overline{\rm DR}$ scheme natural for supersymmetry,
in which every one-loop determinant---gauge, scalar, fermion, and ghost---carries
the universal subtraction constant $3/2$; this value is used here and in all
expressions below.} We note that the
second term above is not a separate
physical longitudinal particle: it is the scalar-like determinant from the
gauge-fixed longitudinal/time-like sector of $A_\mu$, a feature of the $R_{\xi_{\rm gauge}}$
gauge.
The gauge fixing terms requires addition of  Faddeev-Popov ghost terms. Thus
 under an infinitesimal gauge transformation,
 $ \delta A_\mu=\partial_\mu \alpha$, $\delta G=\mx\alpha+\cdots$, where the omitted terms involve fluctuations.  Thus
\begin{equation}
  \delta F=\left(\Box-\xi_{\rm gauge}\mx^2\right)\alpha+\cdots.
\end{equation}
The quadratic ghost Lagrangian is
\begin{equation}
  {
  \lag_{\rm gh}=\bar c\left(-\Box-\xi_{\rm gauge}\mx^2\right)c+\text{interactions}.
  }
\end{equation}
which requires addition of  the ghost mass
\begin{equation}
  {m_{\rm gh}^2(\phi_c)=\xi_{\rm gauge}\mx^2(\phi_c).}
\end{equation}
For an Abelian theory the ghosts decouple from physical scattering, but their
field-dependent determinant still appears in the $R_{\xi_{\rm gauge}}$ effective potential.  Their
contribution is given by
\begin{equation}
  V_{\rm gh}^{(1)}(\phi_c)=
  -\frac{2}{64\pi^2}
  \left(\xi_{\rm gauge}\mx^2\right)^2
  \left[\ln\frac{\xi_{\rm gauge}\mx^2}{\mu_R^2}-\frac32\right],
\end{equation}
where the factor $-2$ reflects the two real Grassmann degrees of freedom of the
complex ghost field.

Next we discuss the contribution of the Goldstone mass in the one-loop determinant.
Suppose $m_{G,0}^2(\phi_c)$ denotes the scalar curvature mass of the would-be Goldstone
before adding the gauge-fixing contribution.  For a potential depending only on the
radial $D$-flat invariant, it is useful to write schematically
\begin{equation}
  m_{G,0}^2(\phi_c)=\frac{1}{\phi_c}\frac{dV_{\rm tree}}{d\phi_c},
\end{equation}
with the present normalization $\Phi=\bar\Phi=\phi_c/2$.  Thus the Goldstone
mass entering the tree-level one-loop determinant is
\begin{equation}
  {
  m_G^2(\phi_c,\xi_{\rm gauge})=m_{G,0}^2(\phi_c)+\xi_{\rm gauge}\mx^2(\phi_c).
  }
\end{equation}
After loop corrections, the corresponding curvature relation is more properly
written as
\begin{equation}
  m_{G,{\rm eff}}^2(\phi_c,\xi_{\rm gauge})=
  \frac{1}{\phi_c}\frac{dV_{\rm eff}}{d\phi_c}+\xi_{\rm gauge}\mx^2(\phi_c).
\end{equation}
At a loop-corrected stationary point $\phi_c=\phi_0$ in Landau gauge, $\xi_{\rm gauge}=0$, one has
\begin{equation}
  {
  m_{G,{\rm eff}}^2(\phi_0,0)=0.
  }
\end{equation}
Away from the stationary
point, the field-dependent Goldstone curvature need not vanish. Collecting the $\xi_{\rm gauge}$-dependent part of the Coleman-Weinberg potential including the unphysical gauge mode, the Goldstone, and the ghosts gives
\begin{equation}
 {
  \begin{aligned}
  V_{\xi_{\rm gauge}}^{(1)}(\phi_c)=\frac{1}{64\pi^2}
  \bigg\{&
  \left[m_{G,0}^2(\phi_c)+\xi_{\rm gauge}\mx^2(\phi_c)\right]^2
  \left[\ln\frac{m_{G,0}^2(\phi_c)+\xi_{\rm gauge}\mx^2(\phi_c)}{\mu_R^2}-\frac32\right]
  \\
  &+\left[\xi_{\rm gauge}\mx^2(\phi_c)\right]^2
  \left[\ln\frac{\xi_{\rm gauge}\mx^2(\phi_c)}{\mu_R^2}-\frac32\right]
  \\
  &-2\left[\xi_{\rm gauge}\mx^2(\phi_c)\right]^2
  \left[\ln\frac{\xi_{\rm gauge}\mx^2(\phi_c)}{\mu_R^2}-\frac32\right]
  \bigg\}.
  \end{aligned}}
\end{equation}
Equivalently,
\begin{align}
  V_{\xi_{\rm gauge}}^{(1)}(\phi_c)=\frac{1}{64\pi^2}
  \bigg\{&
  \left[m_{G,0}^2(\phi_c)+\xi_{\rm gauge}\mx^2(\phi_c)\right]^2
  \left[\ln\frac{m_{G,0}^2(\phi_c)+\xi_{\rm gauge}\mx^2(\phi_c)}{\mu_R^2}-\frac32\right]
  \nonumber\\
  &-\left[\xi_{\rm gauge}\mx^2(\phi_c)\right]^2
  \left[\ln\frac{\xi_{\rm gauge}\mx^2(\phi_c)}{\mu_R^2}-\frac32\right]
  \bigg\}.
\end{align}
The first term is the would-be Goldstone determinant, the second term is the net
combination of the unphysical gauge determinant and the complex ghost determinant.
It is useful to count the degrees of freedom from the $\xi_{\rm gauge}$ dependent part. Here we have
\begin{center}
\begin{tabular}{c|c|c|c}
sector & statistics & mass squared & $(-1)^{2s_i}g_i$ \\
\hline
physical vector & bosonic & $\mx^2$ & $+3$ \\
unphysical gauge scalar & bosonic & $\xi_{\rm gauge}\mx^2$ & $+1$ \\
Goldstone scalar & bosonic & $m_{G,0}^2+\xi_{\rm gauge}\mx^2$ & $+1$ \\
complex ghosts & fermionic Grassmann & $\xi_{\rm gauge}\mx^2$ & $-2$ \\
\end{tabular}
\end{center}

\subsection{Full analysis}
We give now the full CW potential.
The full set of  field-dependent masses entering the one-loop Coleman--Weinberg determinant are as follows:
The physical massive gauge boson has
\begin{equation}
  m_A^2(\phi_c)=g_x^2\phi_c^2,\qquad g_A=3.
\end{equation}
The extra unphysical gauge determinant has
\begin{equation}
  m_{\rm unphys}^2(\phi_c)=\xi_{\rm gauge} g_x^2\phi_c^2,\qquad g_{\rm unphys}=1.
\end{equation}
This is not the physical longitudinal polarization.  The longitudinal polarization is already included in the three physical polarizations of the massive vector.  The Faddeev--Popov ghosts form one complex ghost pair with
\begin{equation}
  m_c^2(\phi_c)=\xi_{\rm gauge} g_x^2\phi_c^2,\qquad g_c=2.
\end{equation}
\nw{The real scalar tree-level masses are those of Table~\ref{tab:masses},
\begin{align}
  m_{\sigma}^2(\phi_c)&=g_x^2\phi_c^2+m_0^2,\\
  m_{h}^2(\phi_c)&=3\lambda_h\phi_c^2-m^2,\\
  m_{G_h^0}^2(\phi_c)&=\lambda_h\phi_c^2-m^2,\\
  m_G^2(\phi_c,\xi_{\rm gauge})&=\lambda_h\phi_c^2-m^2+\xi_{\rm gauge} g_x^2\phi_c^2,
\end{align}
where $\sigma$ is the D-flatness-departing scalar (carrying the soft
mass $m_0^2$), $h$ the radial Higgs, $G_h^0$ the physical
pseudoscalar, and $G$ the would-be Goldstone with tree curvature
$m_{G,0}^2=\lambda_h\phi_c^2-m^2$.  In the flat-direction limit $\lambda_S=0$
($\lambda_h=0$, $m^2=-m_0^2$) the three scalars $h$, $G_h^0$, $G$ become
degenerate at $m_0^2$.}
The last expression should be interpreted as the tree-level mass used inside the one-loop determinant.  At the loop-corrected vacuum the physical Goldstone theorem is instead expressed by $m_{G,{\rm eff}}^2(\phi_0,0)=0$.

\subsection*{Fermionic sector}

The charged higgsino combination
\begin{equation}
  \tilde H_-\equiv \frac{\tilde\Phi-\tilde\Phib}{\sqrt2}
\end{equation}
mixes with the $U(1)_X$ gaugino.  The orthogonal combination
\begin{equation}
  \tilde H_+\equiv \frac{\tilde\Phi+\tilde\Phib}{\sqrt2}
\end{equation}
is massless in the absence of a superpotential coupling.  In the basis $(\tilde\lambda_X,\tilde H_-)$,
\begin{equation}
  \mathcal{M}_F(\phi_c)=
  \begin{pmatrix}
    M_{\tilde\lambda} & \MX(\phi_c)\\
    \MX(\phi_c) & 0
  \end{pmatrix},
  \qquad \MX(\phi_c)=g_x\phi_c.
\end{equation}
The two Majorana masses are
\begin{equation}
  m_{\pm}(\phi_c)=\frac12\left[\sqrt{M_{\tilde\lambda}^2+4\MX^2(\phi_c)}\pm M_{\tilde\lambda}\right],
\end{equation}
or equivalently
\begin{equation}
 {
  m_{\pm}^2(\phi_c)=\MX^2(\phi_c)+\frac12M_{\tilde\lambda}^2
  \pm \frac12M_{\tilde\lambda}\sqrt{M_{\tilde\lambda}^2+4\MX^2(\phi_c)}.}
\end{equation}
These are the two Majorana eigenstates $\tilde\chi_\pm$ of
Table~\ref{tab:masses} and Eq.~(\ref{eq:mgaugino}).
Each Majorana fermion has unsigned multiplicity $g=2$ and enters the one-loop determinant with the fermionic sign $(-1)^{2s}=-1$.  The massless $\tilde H_+$ and the spectator $\tilde S$ do not give field-dependent $m^4\ln m^2$ terms.

\subsection{Degree-of-freedom counting for general $\xi_{\rm gauge}$}

The physical counting in the $\Phi,\Phib$ and $U(1)_X$ sector after Higgsing is
\begin{equation}
  n_B^{\rm phys}=3+3=6,
  \qquad
  n_F^{\rm phys}=4+2=6.
\end{equation}
Here the bosons are the three polarizations of the massive vector plus three physical real scalars.  The fermions are the two Majorana fermions obtained from $(\tilde\lambda_X,\tilde H_-)$ plus the massless Weyl fermion $\tilde H_+$.

In the gauge-fixed $R_{\xi_{\rm gauge}}$ functional determinant the same equality is realized as
\begin{equation}
  n_B^{R_{\xi_{\rm gauge}}}=3+1+4-2=6,
  \qquad
  n_F^{R_{\xi_{\rm gauge}}}=(-2)+(-2)+(-2)=-6.
\end{equation}
with the signed (boson minus fermion) counting.  More explicitly,
\begin{center}
\begin{tabular}{c|c|c}
sector & $(-1)^{2s_i}g_i$ & comment \\
\hline
physical vector $A_\mu$ & $3$ & mass $\MX^2$ \\
unphysical gauge determinant & $1$ & mass $\xi_{\rm gauge}\MX^2$ \\
real scalars $\sigma,h,G_h^0,G$ & $4$ & $G$ has $m_{G,0}^2+\xi_{\rm gauge}\MX^2$ \\
complex ghosts & $-2$ & mass $\xi_{\rm gauge}\MX^2$ \\
\hline
net bosonic/ghost determinant & $6$ & equals physical bosonic count \\
Majorana fermions $\tilde\chi_\pm$ & $-4$ & two Majorana fermions \\
massless Weyl fermion $\tilde H_+$ & $-2$ & field-independent \\
\hline
net fermionic determinant & $-6$ & equals physical fermionic count
\end{tabular}
\end{center}
If the spectator chiral multiplet $S$ is included, it adds two real bosonic degrees of freedom and one Weyl fermion. \nr{However,} since $S$ is decoupled, its contribution is independent of $\phi_c$ and does not affect radiative breaking.
The one-loop potential is
\begin{equation}
  V_{\rm CW}(\phi_c)=\frac{1}{64\pi^2}
  \sum_i (-1)^{2s_i} g_i\, m_i^4(\phi_c)
  \left[\ln\frac{m_i^2(\phi_c)}{\mu_R^2}-\mathcal{C}_i\right].
\end{equation}
\nr{For $\overline{\rm DR}$, $\mathcal{C}_i=3/2$ for all \nr{contributions}
and  the potential in the $R_{\xi_{\rm gauge}}$-gauge is}
\begin{align}
  V_{\rm CW}^{R_{\xi_{\rm gauge}}}(\phi_c)=\frac{1}{64\pi^2}\Bigg\{&
  3\MX^4\left[\ln\frac{\MX^2}{\mu_R^2}-\nw{\frac32}\right]
  +\left(\xi_{\rm gauge}\MX^2\right)^2
  \left[\ln\frac{\xi_{\rm gauge}\MX^2}{\mu_R^2}-\frac32\right]
  \nonumber\\
  &{+m_\sigma^4\left[\ln\frac{m_\sigma^2}{\mu_R^2}-\frac32\right]
  +m_h^4\left[\ln\frac{m_h^2}{\mu_R^2}-\frac32\right]
  +m_{G_h^0}^4\left[\ln\frac{m_{G_h^0}^2}{\mu_R^2}-\frac32\right]}
  \nonumber\\
  &{+m_G^4\left[\ln\frac{m_G^2}{\mu_R^2}-\frac32\right]}
  -2\left(\xi_{\rm gauge}\MX^2\right)^2
  \left[\ln\frac{\xi_{\rm gauge}\MX^2}{\mu_R^2}-\frac32\right]
  \nonumber\\
  &-2m_{+}^4
  \left[\ln\frac{m_{+}^2}{\mu_R^2}-\frac32\right]
  -2m_{-}^4
  \left[\ln\frac{m_{-}^2}{\mu_R^2}-\frac32\right]
  \Bigg\},
  \label{eq:VCWfull_app}
\end{align}
where
\begin{equation}
  \MX^2=g_x^2\phi_c^2.
\end{equation}
Eq.~(\ref{eq:VCWfull_app}) is the general-$R_{\xi_{\rm gauge}}$ form of
Eq.~(\ref{eq:VCW_general}).  In the Landau gauge it reduces to
Eq.~(\ref{eq:VCW_landau_app}) below, equivalent to Eq.~(\ref{eq:VCW_landau}),
whose flat-direction ($\lambda_S=0$) limit is the gauge-sector potential of
Eq.~(\ref{eq:VCW_gauge_full}).
The three explicitly $\xi_{\rm gauge}$-dependent entries are

\begin{equation}
  \begin{split}
  V_{\xi_{\rm gauge}}=\frac{1}{64\pi^2}\Big\{&
  (\xi_{\rm gauge}\MX^2)^2L(\xi_{\rm gauge}\MX^2)\\
  &+(m_{G,0}^2+\xi_{\rm gauge}\MX^2)^2L(m_{G,0}^2+\xi_{\rm gauge}\MX^2)\\
  &-2(\xi_{\rm gauge}\MX^2)^2L(\xi_{\rm gauge}\MX^2)
  \Big\},
  \end{split}
\end{equation}

with
\begin{equation}
  L(x)=\ln\frac{x}{\mu_R^2}-\frac32.
\end{equation}
Thus
\begin{equation}
  V_{\xi_{\rm gauge}}=\frac{1}{64\pi^2}\left\{
  {(m_{G,0}^2+\xi_{\rm gauge}\MX^2)^2L(m_{G,0}^2+\xi_{\rm gauge}\MX^2)}
  -(\xi_{\rm gauge}\MX^2)^2L(\xi_{\rm gauge}\MX^2)
  \right\}.
\end{equation}
\nr{In the Landau gauge, $\xi_{\rm gauge}=0$, and the terms proportional to $(\xi_{\rm gauge}\MX^2)^2\ln(\xi_{\rm gauge}\MX^2)$ vanish in the limit $x^2\ln x\to0$, \nw{while the Goldstone contribution becomes $m_{G,0}^4[\ln(m_{G,0}^2/\mu_R^2)-3/2]$ with $m_{G,0}^2=\lambda_h\phi_c^2-m^2$.}
Thus the full one-loop Coleman--Weinberg potential along the $D$-flat 
0direction is given by}
\begin{align}
  V_{\rm CW}(\phi_c)=\frac{1}{64\pi^2}\Big\{&
  3\MX^4\,L(\MX^2)
  {+m_\sigma^4\,L(m_\sigma^2)
  +m_h^4\,L(m_h^2)
  +2\,m_{G,0}^4\,L(m_{G,0}^2)}
  \nonumber\\
  &-2m_{+}^4\,L(m_{+}^2)
  -2m_{-}^4\,L(m_{-}^2)
  \Big\},
  \label{eq:VCW_landau_app}
\end{align}
\nw{with $L(x)=\ln(x/\mu_R^2)-3/2$, $m_\sigma^2=\MX^2+m_0^2$,
$m_h^2=3\lambda_h\phi_c^2-m^2$, and $m_{G,0}^2=\lambda_h\phi_c^2-m^2$; the factor
$2$ in the $m_{G,0}^4$ term counts the degenerate pseudoscalar $G_h^0$ and
would-be Goldstone $G$.  The full effective potential along the flat direction
is then $V_{0}(\phi_c)=V_{\rm tree}(\phi_c)+V_{\rm CW}(\phi_c)$, with
$V_{\rm tree}$ the tree potential of Eq.~(\ref{eq:Vtree_flat}).}

\subsection{Curvature at the origin and natural supercooling}
\label{sec:app_barrier}

Natural supercooling requires a barrier already in the zero-temperature
potential $V_0=V_{\rm tree}+V_{\rm CW}$; since $V_0$ is even in $\phi_c$ this
reduces to a positive curvature at the origin, \nr{as seen in}
Eq.~(\ref{eq:barrier_condition}). We derive here the curvature \nr{given} in
Eq.~(\ref{eq:V0pp}). Thus 
writing each Coleman--Weinberg term as $f(m_i^2)$ where 
$f(u)=u^2[\ln(u/\mu_R^2)-\tfrac32]$, \nr{we get}
\begin{equation}
  f'(u)=2u\Big(\ln\frac{u}{\mu_R^2}-1\Big),\qquad
  f''(u)=2\ln\frac{u}{\mu_R^2}.
\end{equation}
Since every mass-squared depends on $\phi_c$ only through $\phi_c^2$, one has
$dm_i^2/d\phi_c=0$ at the origin, and the second derivative of each term \nr{reduces}
to
\begin{equation}
  \frac{d^2}{d\phi_c^2}f(m_i^2)\bigg|_{0}
  =f'\!\big(m_i^2(0)\big)\,\frac{d^2m_i^2}{d\phi_c^2}\bigg|_{0}.
\end{equation}
Only fields with nonzero mass at the origin contribute: the gauge boson
($m_A^2=g_x^2\phi_c^2$) and the light gaugino
($m_-^2\simeq g_x^4\phi_c^4/M_{\tilde\lambda}^2$) vanish there, and the scalars
$h,G_h^0,G$ are $\phi_c$-independent for $\lambda_S=0$. The survivors are the
D-scalar $\sigma$ and the heavy gaugino $\tilde\chi_+$, with
\begin{equation}
  m_\sigma^2(0)=m_0^2,\quad
  \frac{d^2m_\sigma^2}{d\phi_c^2}\bigg|_0=2g_x^2;
  \qquad
  m_+^2(0)=M_{\tilde\lambda}^2,\quad
  \frac{d^2m_+^2}{d\phi_c^2}\bigg|_0=4g_x^2,
\end{equation}
the gaugino values following from Eq.~(\ref{eq:mgaugino}). With multiplicities
$g_\sigma=1$, $g_{\tilde\chi_+}=2$ (the gaugino entering with fermionic sign $(-1)^{2s}=-1$), and the tree term $V_{\rm tree}''(0)=m_0^2$,
\begin{align}
  \left.\frac{d^2V_0}{d\phi_c^2}\right|_0
  &=m_0^2+\frac{1}{64\pi^2}\Big[
  2m_0^2\big(\ln\tfrac{m_0^2}{\mu_R^2}-1\big)\,2g_x^2
  -2\cdot 2M_{\tilde\lambda}^2\big(\ln\tfrac{M_{\tilde\lambda}^2}{\mu_R^2}-1\big)\,4g_x^2
  \Big] \nonumber\\
  &=m_0^2+\frac{g_x^2}{16\pi^2}\!\left[
  m_0^2\Big(\ln\frac{m_0^2}{\mu_R^2}-1\Big)
  -4M_{\tilde\lambda}^2\Big(\ln\frac{M_{\tilde\lambda}^2}{\mu_R^2}-1\Big)\right],
\end{align}
which is Eq.~(\ref{eq:V0pp}). The one-loop piece is suppressed by
$g_x^2/16\pi^2$, so $d^2V_0/d\phi_c^2|_0\simeq m_0^2>0$: the soft scalar mass
keeps the origin a local minimum and the zero-temperature barrier of natural
supercooling forms automatically.

\subsection{Stability of the radiative vacuum}
\label{sec:app_stability}

The radiative minimum is not automatically a stable vacuum; its stability is set
by the curvature of the full $T=0$ potential $V_0=V_{\rm tree}+V_{\rm CW}$ at the
symmetry-breaking point $\phi_0$. From the local expansion
Eq.~(\ref{eq:Veff_expand}), $\phi_0$ is a genuine minimum if and only if it is a
stationary point, $V_{\rm eff}'(\phi_0)=0$, with positive radial mass
$m_\rho^2\equiv V_{\rm eff}''(\phi_0)>0$. We now derive this curvature exactly.

The curvature follows by differentiating the one-loop potential twice with
respect to $\phi_c$. Writing the Coleman--Weinberg sum as
\begin{equation}
  V_{\rm CW}(\phi_c)=\frac{1}{64\pi^2}\sum_i(-1)^{2s_i}g_i\,m_i^4(\phi_c)
  \left[\ln\frac{m_i^2(\phi_c)}{\mu_R^2}-\tfrac32\right]
  \equiv\frac{1}{64\pi^2}\sum_i(-1)^{2s_i}g_i\,f\!\big(m_i^2(\phi_c)\big),
\end{equation}
the whole $\phi_c$-dependence of the $i$-th term is carried by the
field-dependent mass-squared $u\equiv m_i^2(\phi_c)$ through the single function
$f(u)=u^2[\ln(u/\mu_R^2)-\tfrac32]$, whose derivatives are
\begin{align}
  f'(u)&=2u\ln\frac{u}{\mu_R^2}+u^2\cdot\frac1u-3u
        =2u\Big(\ln\frac{u}{\mu_R^2}-1\Big),
  \label{eq:fprime}\\
  f''(u)&=2\ln\frac{u}{\mu_R^2}+2u\cdot\frac1u-2
        =2\ln\frac{u}{\mu_R^2}.
  \label{eq:fpprime}
\end{align}
Since $\phi_c$ enters only through $u=m_i^2(\phi_c)$, the chain rule gives, with
$m_i^{2\prime}\equiv dm_i^2/d\phi_c$ and $m_i^{2\prime\prime}\equiv
d^2m_i^2/d\phi_c^2$,
\begin{align}
  V_{\rm CW}'(\phi_c)&=\frac{1}{64\pi^2}\sum_i(-1)^{2s_i}g_i\,
    f'(m_i^2)\,m_i^{2\prime},\\
  V_{\rm CW}''(\phi_c)&=\frac{1}{64\pi^2}\sum_i(-1)^{2s_i}g_i
    \Big[f''(m_i^2)\,(m_i^{2\prime})^2+f'(m_i^2)\,m_i^{2\prime\prime}\Big].
\end{align}
Inserting $f'$ and $f''$ from Eqs.~(\ref{eq:fprime})--(\ref{eq:fpprime}) and
factoring out the common $2$,
\begin{equation}
  V_{\rm CW}''(\phi_c)=\frac{1}{32\pi^2}\sum_i(-1)^{2s_i}g_i\!\left[
  \ln\frac{m_i^2}{\mu_R^2}\Big(\frac{dm_i^2}{d\phi_c}\Big)^{\!2}
  +m_i^2\Big(\ln\frac{m_i^2}{\mu_R^2}-1\Big)\frac{d^2m_i^2}{d\phi_c^2}\right].
  \label{eq:VCWpp_app}
\end{equation}
Adding the tree term $V_{\rm tree}''=m_0^2$, the exact radial mass is
\begin{equation}
  m_\rho^2=m_0^2+\frac{1}{32\pi^2}\sum_i(-1)^{2s_i}g_i\!\left[
  \Big(\frac{dm_i^2}{d\phi_c}\Big)^{\!2}\ln\frac{m_i^2}{\mu_R^2}
  +m_i^2\frac{d^2m_i^2}{d\phi_c^2}\Big(\ln\frac{m_i^2}{\mu_R^2}-1\Big)
  \right]_{\phi_0}.
\end{equation}
Stability of the radiative vacuum is the
requirement that this curvature be positive,
\begin{equation}
  m_\rho^2=V_{\rm eff}''(\phi_0)>0,
\end{equation}
imposed on the full spectrum of Table~\ref{tab:masses} throughout the parameter
scan. 

\section{Phase Transition and GW Spectrum Formulas}
\label{sec:appendix_PT}

We collect the standard formulas for the phase transition dynamics
and GW spectrum computation used in this
work~\cite{Athron:2023xlk,Li:2025nja,Caprini:2015zlo,Grojean:2006bp,Weir:2017wfa,Caprini:2018mtu,Mazumdar:2018dfl,Hindmarsh:2020hop,Schmitz:2020syl}.

\subsection{Phase transition dynamics}
We collect the nucleation formalism and define the mean bubble
separation $\Rstar$, the strength $\alpha$, and the inverse
duration $\beta/H$ used in the GW computation.

The bubble nucleation rate per unit volume
is~\cite{Coleman:1977py,Callan:1977pt,Linde:1980tt,Turner:1992tz,Megevand:2016lpr}
\begin{align}
    \Gamma(T) \simeq T^4
    \left(\frac{S_3(T)}{2\pi T}\right)^{3/2}
    \exp\left(-S_3(T)/T\right),
    \label{eq:Gamma_app}
\end{align}
with $S_3(T)$ the bounce action. The fraction of space still in
the false vacuum is
\begin{align}
    P_f(T) = e^{-I(T)}\,,
    \label{eq:Pf_app}
\end{align}
governed by the nucleation integral
\begin{align}
    I(T) = \frac{4\pi}{3}\,v_w^3
    \int_{T}^{\Tc} \frac{\Gamma(T')}{T'^{\,4}\, H(T')}
    \left(\int_{T}^{T'} \frac{dT''}{H(T'')}\right)^3 dT'\,,
    \label{eq:I_app}
\end{align}
where $v_w$ is the bubble-wall velocity and $H(T)$ the Hubble
rate. The percolation temperature $\Tp$ is fixed by
$P_f(\Tp) = 0.71$.

The scale that controls the GW spectrum is the mean bubble
separation at percolation,
\begin{align}
    \Rstar = \left(T^3\int_{T}^{\Tc}
    \frac{\Gamma(T')P_f(T')}{T'^4 H(T')}dT'\right)^{-1/3},
    \label{eq:Rstar_app}
\end{align}
in terms of which we define the inverse duration
\begin{align}
    \frac{\beta}{H} \equiv \frac{(8\pi)^{1/3}}{H_*\Rstar}\,,
    \qquad H_* \equiv H(\Tp)\,.
    \label{eq:betaH_def}
\end{align}
We adopt $\Rstar$ rather than the conventional single-point
estimate $\beta/H = T\,d(S_3/T)/dT|_{\Tp}$, which is inaccurate for
the strongly supercooled transitions studied here, where the
nucleation window is narrow and $\Gamma$ sharply peaked; the
integral~(\ref{eq:Rstar_app}) remains robust~\cite{Li:2025nja}.

The transition strength is
\begin{align}
    \alpha = \frac{\Delta\bar\theta(\Tp)}{\rho_{\rm rad}(\Tp)}\,,
    \qquad
    \Delta\bar\theta = \Delta V - \frac{T}{4}\frac{d\Delta V}{dT}\,,
    \label{eq:alpha_app}
\end{align}
with $\Delta V = V_{\rm eff}(\phi_f, T) - V_{\rm eff}(\phi_t, T)$
the difference in potential energy between the false and true vacua
and $\rho_{\rm rad}$ the total radiation energy density.

\paragraph{Percolation criterion.}
For the transition to complete, the physical false-vacuum volume
$\mathcal{V}_f(t) = P_f(t)\,(a(t)/a_0)^3$ must be decreasing at
percolation~\cite{Athron:2022mmm,Athron:2023xlk},
\begin{align}
    \frac{1}{P_f}\frac{dP_f}{dt}\bigg|_{\Tp} + 3H(\Tp) < 0\,.
    \label{eq:perc_completion}
\end{align}
With $P_f = e^{-I}$ and $dt = -dT/(HT)$ in the radiation era, this
is equivalently
\begin{align}
    T\,\frac{dI}{dT}\bigg|_{\Tp} > 3\,.
    \label{eq:perc_TdIdT}
\end{align}
For the narrow nucleation window of a strongly supercooled
transition the integral~(\ref{eq:I_app}) is dominated by
$T \simeq \Tp$, where the mean bubble radius is
$\bar r \simeq v_w/H_*$ and
\begin{align}
    I(\Tp) \simeq \frac{4\pi}{3}\,\frac{v_w^3}{(H_*\Rstar)^3}\,.
    \label{eq:I_saddle}
\end{align}
Because the dominant temperature dependence is the exponential
in $\Gamma$ (with $S_3/\Tp \sim 140$ and $I(\Tp) \approx 0.34$),
the criterion reduces, for $v_w \approx 1$, to
\begin{align}
    \frac{(8\pi)^{1/3}}{H_*\Rstar} > 3\,,
    \qquad\text{i.e.}\quad \frac{\beta}{H} > 3\,,
    \label{eq:perc_HRstar}
\end{align}
so the bubbles must be separated by less than a Hubble radius
to overlap and fill space. All benchmarks here satisfy this; the
strongest ($M_{\tilde\lambda} = 0.23\,v_X$) lies close to the
boundary, while $M_{\tilde\lambda} \geq 0.235\,v_X$ fails to
percolate (Table~\ref{tab:benchmarks}).

\subsection{GW power spectrum}
\label{sec:appendix_gw_spectrum}

The GW power spectrum receives contributions from bubble collisions
(scalar field), sound waves, and
turbulence~\cite{Caprini:2015zlo,Caprini:2019egz,Athron:2023xlk,Giese:2020znk,Apreda:2001us,Maggiore:1999vm}:
$\Omega_{\rm GW} = \Omega_{\rm col} + \Omega_{\rm sw}
+ \Omega_{\rm turb}$.
The redshifted spectrum today is
\begin{align}
    \Omega_{\rm GW}^0(f_0) = \mathcal{R}_\Omega\,
    \Omega_{\rm GW}\left(\frac{f_0}{\mathcal{R}_f}\right),
\end{align}
with redshift factors
\begin{align}
    \mathcal{R}_f &= \left(\frac{h_{\rm eff}^0}
    {h_{\rm eff}}\right)^{1/3}
    \frac{T_0}{T_{v,*}}\,, \qquad
    \mathcal{R}_\Omega \simeq 2.473\times 10^{-5}h^{-2}
    \left(\frac{h_{\rm eff}^0}
    {h_{\rm eff}}\right)^{4/3}
    \left(\frac{g_{\rm eff}}{2}\right).
\end{align}
The redshift is referenced to the \emph{visible}-sector temperature at the
transition, $T_{v,*}=\Tp$. This is the relevant scale because the latent heat
reheats the \emph{hidden} sector and not the visible one---the portal transfer
is feeble ($\propto\delta^2$)---so the visible temperature that sets the scale
factor does not reheat. The reheating enters instead through the hidden
contribution to the effective degrees of freedom, evaluated at the reheated
true-vacuum temperature $T_{h,t}=\xi_t\,T_{v,*}$ and weighted by the reheated
ratio $\xi_t$,
\begin{align}
    g_{\rm eff} &= g_{\rm eff}^v(T_{v,*}) + g_{\rm eff}^h(T_{h,t})\,\xi_t^4\,,
    &
    h_{\rm eff} &= h_{\rm eff}^v(T_{v,*}) + h_{\rm eff}^h(T_{h,t})\,\xi_t^3\,.
\end{align}
Here $\xi_t$ is the true-vacuum (reheated-interior) ratio of the coupled
system~(Section~\ref{sec:full_system}); the percolation temperature $\Tp$ and
the false-vacuum ratio $\xi_f$, which control the source amplitude and $\Rstar$,
are distinct from the reheated $\xi_t$ that sets the redshifted spectrum.
We give below the formulae for the power spectrum from bubble collision,
sound waves, and turbulence.

\paragraph{Bubble collision (envelope approximation)~\cite{Kosowsky:1992rz,Kosowsky:1992vn,Caprini:2007xq,Huber:2008hg,Cutting:2018tjt}:}
\begin{align}
    \Omega_{\rm col}(f) = 1.67\times 10^{-5}
    \left(\frac{H_*\Rstar}{1}\right)^2
    \left(\frac{\kappa_\phi\,\alpha}{1+\alpha}\right)^2
    \left(\frac{100}{g_*}\right)^{1/3}
    \frac{3.8\,(f/f_{\rm col})^{2.8}}
    {1 + 2.8\,(f/f_{\rm col})^{3.8}}\,,
\end{align}
where $\kappa_\phi$ is the fraction of vacuum energy converted to
scalar field gradient energy and the peak frequency is
$f_{\rm col} \simeq 1.65\times 10^{-5}\;\text{Hz}\,
(0.62/(1.8 - 0.1\,v_w + v_w^2))\,
(T_*/100\;\gev)\,(1/(H_*\Rstar))\,(g_*/100)^{1/6}$.

\paragraph{Sound waves~\cite{Hindmarsh:2013xza,Hindmarsh:2015qta,Hindmarsh:2017gnf,Hindmarsh:2016lnk,Hindmarsh:2019phv,Ellis:2020awk}:}
\begin{align}
    \Omega_{\rm sw}(f) = 2.65\times 10^{-6}
    \left(\frac{H_*\Rstar}{1}\right)
    \left(\frac{\kappa_f\,\alpha}{1+\alpha}\right)^2
    \left(\frac{100}{g_*}\right)^{1/3}\,v_w\,
    \left(\frac{f}{f_{\rm sw}}\right)^3
    \left(\frac{7}{4 + 3\,(f/f_{\rm sw})^2}\right)^{7/2},
\end{align}
where $\kappa_f$ is the fluid kinetic energy fraction and
$f_{\rm sw} \simeq 1.9\times 10^{-5}\;\text{Hz}\,
(1/v_w)\,(T_*/100\;\gev)\,\\ 
(1/(H_*\Rstar))\,
(g_*/100)^{1/6}$.

\paragraph{Turbulence~\cite{Kosowsky:2001xp,Caprini:2009yp,RoperPol:2019wvy}:}
\begin{align}
    \Omega_{\rm turb}(f) = 3.35\times 10^{-4}
    \left(\frac{H_*\Rstar}{1}\right)
    \left(\frac{\epsilon\,\kappa_f\,\alpha}{1+\alpha}\right)^{3/2}
    \left(\frac{100}{g_*}\right)^{1/3}\,v_w\,
    \frac{(f/f_{\rm turb})^3}
    {(1 + f/f_{\rm turb})^{11/3}\,
    (1 + 8\pi f/h_*)}\,,
\end{align}
where $\epsilon \approx 0.05$--$0.10$ is the fraction of bulk
kinetic energy converted to turbulence,
$f_{\rm turb} \simeq 2.7\times 10^{-5}\;\text{Hz}\,
(1/v_w)\,(T_*/100\;\gev)\,(1/(H_*\Rstar))\,(g_*/100)^{1/6}$,
and $h_* = 1.65\times 10^{-5}\;\text{Hz}\,
(T_*/100\;\gev)\,\\
(g_*/100)^{1/6}$ is the Hubble rate at
production redshifted to today.

\paragraph{Peak frequency:}
The peak frequency after redshift scales as
\begin{align}
    f_0 \approx 1.65\times 10^{-5}\;\text{Hz} \times
    \frac{T_*}{100\;\gev} \times \frac{1}{H_*\Rstar}
    \times \left(\frac{g_*}{100}\right)^{1/6},
    \label{eq:fpeak}
\end{align}
where, consistently with the redshift factors above, $T_*=T_{v,*}=\Tp$ is the
visible percolation temperature (the visible sector does not reheat) and
$g_*=g_{\rm eff}$ is the reheated effective number of degrees of freedom,
including the hidden contribution at $T_{h,t}=\xi_t T_{v,*}$.

\paragraph{Wall velocity and efficiency factors.}
The partition of the released vacuum energy between the bubble walls and the
hidden-sector plasma---and hence the relative weight of the three GW sources
above---is fixed self-consistently from the hidden-sector hydrodynamics in three
steps, all evaluated at the percolation point.

\emph{(i) Sound speeds.} The squared sound speeds in the symmetric and broken
phases, $c_{s,s}^2$ and $c_{s,b}^2$, are computed from the hidden-sector
enthalpy and pressure derived from $V_{\rm eff}$,
$c_s^2 = (dp/dT)/(d\rho/dT)$~\cite{Giese:2020rtr,Giese:2020znk}; near the strongly
supercooled transition they depart from the relativistic value $1/3$.

\emph{(ii) Wall velocity.} The terminal wall velocity $v_w$ is obtained by
solving the hydrodynamic matching conditions across the wall in the
enthalpy-ratio ($\nu$--$\mu$) formulation~\cite{Ai:2023see,Laurent:2022jrs},
using the Chapman--Jouguet velocity
\begin{equation}
  v_J = \frac{\sqrt{c_{s,b}^2}\left[1+\sqrt{3\,\alpha_{\rm DS}\,
  (1-c_{s,b}^2+3\,c_{s,b}^2\,\alpha_{\rm DS})}\right]}
  {1+3\,c_{s,b}^2\,\alpha_{\rm DS}}
  \label{eq:vJ}
\end{equation}
as the deflagration--detonation boundary. The wall hydrodynamics are governed by
the \emph{hidden-sector} strength $\alpha_{\rm DS}=\Delta\bar\theta/\rho_{\rm rad}^h$
(not the visible-diluted $\atot$), since the wall propagates in the hidden plasma.
For the deeply supercooled benchmarks the matching has no deflagration/hybrid
solution and the wall is a detonation, $v_w\to1$.

\emph{(iii) Efficiency factors and energy budget.} The fraction of vacuum energy
converted to bulk fluid motion is the bag-model efficiency
$\kappa_f=\kappa(\alpha_{\rm DS},v_w,c_{s,s}^2,c_{s,b}^2)$ of the $\nu$--$\mu$
model~\cite{Espinosa:2010hh,Giese:2020rtr}, obtained by integrating the
self-similar fluid profile (again at the hidden-sector strength
$\alpha_{\rm DS}$). Whether the wall runs away is decided by comparing
the transition strength to the Bödeker--Moore leading-order maximal
friction~\cite{Bodeker:2009qy,Bodeker:2017cim}, expressed as a strength
\begin{equation}
  \alpha_\infty =
  \frac{T_{h,f}^2}{24\,\rho_{\rm rad}}
  \sum_i c_i\,n_i\,\Delta m_i^2,
  \label{eq:alpha_inf}
\end{equation}
where $\rho_{\rm rad}$ is the total radiation density,
$\Delta m_i^2=m_i^2(\phi_b)-m_i^2(0)$ is the field-dependent mass-squared jump
across the wall, $n_i$ counts the internal degrees of freedom, and $c_i$ is the
thermal weight of species $i$ in the leading-order friction term.  In the
present benchmarks the massive hidden gauge boson gives the dominant
contribution. The kinetic energy is then partitioned---comparing
$\alpha_\infty$ to the
\emph{total} strength $\atot$, both normalized to $\rho_{\rm rad}$ (the ratio
$\alpha_\infty/\atot$ being independent of this choice)---as
\begin{equation}
  (\kappa_\phi,\,\kappa_{\rm sw}) =
  \begin{cases}
    (0,\;\kappa_f), & \atot<\alpha_\infty \quad\text{(terminal wall)},\\[4pt]
    \Big(1-\dfrac{\alpha_\infty}{\atot},\;
      \dfrac{\alpha_\infty}{\atot}\,\kappa_\infty\Big), & \atot>\alpha_\infty
      \quad\text{(runaway)},
  \end{cases}
  \label{eq:kappa_partition}
\end{equation}
where $\kappa_\phi$ is the scalar-gradient (bubble-collision) fraction,
$\kappa_{\rm sw}$ the sound-wave fraction, and $\kappa_\infty$ is the efficiency
evaluated at $\alpha_\infty$; the turbulent fraction is taken to be
$\kappa_{\rm turb}=0.1\,\kappa_{\rm sw}$~\cite{Caprini:2009yp,RoperPol:2019wvy}.
The applicable regime---terminal wall ($\atot<\alpha_\infty$, $\kappa_\phi=0$)
or runaway ($\atot>\alpha_\infty$, $\kappa_\phi>0$)---is thus determined per
benchmark by Eq.~(\ref{eq:kappa_partition}). We note that
Eq.~(\ref{eq:alpha_inf}) is the leading-order Bödeker--Moore friction; in a
gauge theory the next-to-leading transition-radiation friction grows with the
wall Lorentz factor~\cite{Bodeker:2017cim} and can keep the energy in the fluid
(sound waves) even when the leading-order criterion indicates runaway, an effect
not captured by Eq.~(\ref{eq:kappa_partition}).

\section{Cross Sections for Hidden Sector Processes}
\label{sec:appendix_xsec}

This appendix collects the cross sections and decay widths used in the yield
equations of Appendix~\ref{sec:appendix_boltzmann} and in the dark-quark
thermalization analysis of Section~\ref{sec:dq_in_PT}.  The expressions are
kept at the level needed for the cosmological evolution; subleading channels
are retained only where they affect the freeze-out or decay history.

\subsection{Gaugino pair annihilation:
$\tilde\chi\tilde\chi \to \chi\chi$}

The two Majorana mass eigenstates $\tilde\chi_\pm$
(Section~\ref{sec:fermion_dof}) annihilate into hidden Higgs pairs through
$t/u$-channel exchange, with the SUSY gauge-Yukawa vertex
$g_x\,\tilde\chi\,\tilde H\,\chi$.  The mixing angle $\theta$ that diagonalizes
the gaugino-Higgsino mass matrix~(\ref{eq:ferm_mass_matrix}) fixes the relevant
couplings,
$y_{\rm self} = g_x\sin 2\theta/4$ (self-annihilation) and
$y_{\rm co} = g_x\cos 2\theta/2$ (co-annihilation).
For a given initial pair the $s$-wave cross section is
\begin{align}
    \sigma v\big|_{\tilde\chi_a\tilde\chi_b \to \chi\chi}
    = \frac{y_{ab}^4}{8\pi\,m_{ab}^2}\,
    \sqrt{1 - \frac{m_\chi^2}{m_{ab}^2}}\,,
    \label{eq:sv_gaug_hh}
\end{align}
where $m_{ab}$ is the average mass of the initial pair.  The effective thermal
cross section used in the Boltzmann equation is the Boltzmann-weighted sum over
the self- and co-annihilation channels,
\begin{align}
    \langle\sigma v\rangle_{\tilde\chi\tilde\chi\to\chi\chi}^{\rm eff}
    = \sum_{a,b \in \{+,-\}} \frac{w_a\,w_b}{w_{\rm tot}^2}\,
    \sigma v\big|_{ab}\,,
    \quad
    w_\pm = 2\,e^{-m_\pm/T_h}\,,
    \label{eq:sv_gaug_eff}
\end{align}
where $w_{\rm tot}=w_+ + w_-$.

\subsection{Lighter Higgs scalar pair annihilation:
$\sigma\sigma \to \gamma'\gamma'$}

The lighter scalar annihilates dominantly into dark photons.  Including the
contact interaction and the $t/u$-channel diagrams, the $s$-wave cross section
for a real scalar pair into massive vectors is
\begin{align}
    \sigma v\big|_{\sigma\sigma \to \gamma'\gamma'}
    = \frac{g_x^4}{16\pi\,m_{\sigma}^2}\,
    \sqrt{1 - \frac{m_A^2}{m_{\sigma}^2}}\,
    \left(1 + \frac{m_A^2}{2m_{\sigma}^2}\right).
    \label{eq:sv_sigD_gg}
\end{align}

\subsection{Dark quark annihilation channels (symmetric phase)}
\label{sec:appendix_dq_xsec}

In the symmetric phase, $T_h>\Tp$, the dark quark remains in chemical
equilibrium through the following $2\to2$ processes.

\paragraph{Channel 1: $q\bar q \to \gamma'\gamma'$.}
\begin{align}
    \sigma_1(s) = \frac{2\pi\alpha_X^2 q_X^4}{s}\,,
    \qquad
    \langle\sigma v\rangle_1
    \sim \frac{2\pi\alpha_X^2 q_X^4}{T_h^2}\,.
    \label{eq:sigma_qqbar_gamgam}
\end{align}

\paragraph{Channel 2: $q\bar q \to \tilde\lambda_X\tilde\lambda_X$.}
\begin{align}
    \sigma_2(s) = \frac{\pi\alpha_X^2 q_X^4}{s}
    \times\frac{s^2}{(s + 2m_{\tilde q}^2)^2}
    \times\Theta\!\left(\sqrt{s} - 2M_{\tilde\lambda}\right),
    \label{eq:sigma_qqbar_gauginogaugino}
\end{align}
Here $m_{\tilde q}^2 = m_0^2 + q_X^2 D_X$.  Near freeze-out the gaugino
threshold gives a Boltzmann suppression
$\sim e^{-2M_{\tilde\lambda}/T_h} \approx 10^{-4}$.

\paragraph{Channel 3:
$q\bar q \to \tilde H_\Phi\tilde{\bar H}_{\bar\Phi}$.}
\begin{align}
    \sigma_3(s) = \frac{4\pi\alpha_X^2 q_X^2}{3s}\,.
    \label{eq:sigma_qqbar_higgsinohiggsino}
\end{align}

\paragraph{Total interaction rate.}
\begin{align}
    \Gamma_{\rm tot} \sim \frac{3\zeta(3)\alpha_X^2}{2\pi^2}\,T_h
    \times\!\left(2q_X^4 + q_X^4\,e^{-2M_{\tilde\lambda}/T_h}
    + \frac{4q_X^2}{3}\right),
    \label{eq:Gamma_tot}
\end{align}
At the transition this gives
$\Gamma_{\rm tot}/H|_{T_h = \Tp} \approx 3\times10^8$
\label{eq:Gamma_over_H}, confirming that the dark quark remains deep in
chemical equilibrium throughout the symmetric phase.

\subsection{Hidden Higgs decay}
\label{sec:appendix_chi_decay}

In the broken phase the hidden Higgs $\chi$ has no fast two-body decay into a
single dark photon: the covariant derivative couples $\chi$ to gauge-boson
pairs.  The relevant slow channels are the Yukawa decay,
$\Gamma_\chi^{(y)}=y_q^2m_\chi/(8\pi)\sim10^{-21}\;\gev$
($\tau\sim0.5\;$ms), and the gravity-mediated decay,
$\Gamma_\chi^{(\rm grav)}\sim m_\chi^3/\Mpl^2\sim10^{-20}\;\gev$
($\tau\sim10^{-5}\;$s).  The latter dominates, so $\chi$ decays before BBN and
deposits most of its energy into the visible sector.  The subdominant branching
fraction to dark quarks is estimated as
\begin{align}
    \mathrm{BR}(\chi \to q\bar q) \approx
    \frac{1}{1 + \Delta N_{\rm eff}^{\rm SM}},
    \label{eq:BR_chi_qq}
\end{align}
so the direct injection $\Delta Y_q=2\,\mathrm{BR}\,Y_\chi$ is only a
few-percent correction for the benchmarks in Table~\ref{tab:cosmo_constraints}.

\subsection{Dark photon portal processes}

The kinetic-mixing portal allows visible fermion pairs, denoted by $ii$, to
produce dark photons through $ii\to\gamma'$, and allows the inverse decay
$\gamma'\to ii$.  Both rates scale as $\delta^2$; in particular
$\Gamma_{\gamma'\to ii}\propto\delta^2\alpha_{\rm em}m_{\gamma'}$.  For
$\delta\lesssim10^{-7}$ these portal processes are negligible compared with the
hidden-sector interactions.

\section{Complete Boltzmann System: ODE Equations, Particle Yields, and Decay Channels}
\label{sec:appendix_boltzmann}

\subsection{Instantaneous reheating approximation}
\label{sec:instant_reheat_app}

The instantaneous-reheating estimate is useful as a diagnostic, but it is not
the evolution used for the GW prediction.  In this approximation the latent heat
released at the transition is assumed to be deposited at once into a homogeneous
hidden bath,
\begin{equation}
    \rho_{\rm rad}^h(\Treh) \simeq \rho_{\rm rad}^h(\Tp)
    + \rho_{\rm vac}^h(\Tp),
    \label{eq:instant_reheat}
\end{equation}
yielding
\begin{equation}
    \Treh = (1 + \ah)^{1/4}
    \left(\frac{g_{\rm eff}^h(\Tp)}{g_{\rm eff}^h(\Treh)}\right)^{1/4}
    \Tp\,,
    \label{eq:Treh_instant}
\end{equation}
where $\ah = \Delta\bar\theta(\Tp)/\rho_{\rm rad}^h(\Tp)$ is the
hidden-sector transition strength.  Equation~(\ref{eq:Treh_instant}) gives the
completion temperature expected if the portal leakage and Hubble expansion
during percolation are negligible.  This is why it agrees well with the
two-temperature result for the weak-portal rows of
Table~\ref{tab:Treh_compare}.

The approximation has two limitations for the present analysis.  First, the
transition completes over a finite nucleation history rather than by a
discontinuous jump.  Second, and more importantly, the released energy heats the
true-vacuum interior, while the false-vacuum exterior remains cold and controls
the tunneling rate.  The main text therefore uses Eq.~(\ref{eq:Treh_instant})
only as a comparison to the final true-vacuum temperature
$T_{h,t}^{\rm completion}$; it does not use the reheated temperature in
$\Gamma$.

\subsection{Derivation of the temperature-ratio equation}
\label{sec:appendix_xi_derivation}

The coupled thermal evolution follows from energy conservation in an FRW
background.  The visible and hidden sectors exchange energy through the
kinetic-mixing portal, with rates $j_v$ and $j_h$ proportional to
$\delta^2$.  Before the transition, or on the cold false-vacuum branch, the
source-free equations are
\begin{align}
    \frac{d\rho_v}{dt} + 3H(\rho_v + p_v) &= j_v\,,
    \label{eq:drho_v_dt} \\
    \frac{d\rho_h}{dt} + 3H(\rho_h + p_h) &= j_h\,.
    \label{eq:drho_h_dt}
\end{align}
where $H=\dot a/a$.  Each sector is assumed to be in local thermal equilibrium.
It is useful to write the enthalpy terms as
\begin{equation}
    \rho_i+p_i=\frac{4}{3}\,\sigma_i\rho_i\,,
    \qquad
    \sigma_i\equiv \frac{3}{4}\left(1+\frac{p_i}{\rho_i}\right),
\end{equation}
and similarly for the total fluid,
\begin{equation}
    \rho+p=\frac{4}{3}\,\sigma\rho\,,
    \qquad
    \sigma\equiv\frac{3}{4}\left(1+\frac{p_v+p_h}{\rho_v+\rho_h}\right).
\end{equation}
For radiation $\sigma_i=1$, but we keep the notation general because the
effective degrees of freedom vary with temperature.

Using $d\rho_i/dt=(d\rho_i/dT_i)\dot T_i$, the hidden equation gives
\begin{equation}
    \dot T_{h,f}
    = \frac{j_h-4H\sigma_h\rho_h}{d\rho_h/dT_h}\,.
    \label{eq:Tdot_hf_app}
\end{equation}
The visible temperature is obtained from total energy conservation.  Adding the
two source-free equations gives
\begin{equation}
    \dot\rho_v+\dot\rho_h+4H\sigma\rho=0\,,
\end{equation}
after using $j_v+j_h=0$ for energy exchange between the two sectors.  Eliminating
$\dot\rho_h$ with the hidden-sector equation yields
\begin{equation}
    \dot T_v =
    -\frac{4H\sigma\rho-4H\sigma_h\rho_h-j_v}{d\rho_v/dT_v}\,.
    \label{eq:Tdot_v_app}
\end{equation}
Writing $T_{h,f}=\xi_fT_v$ for the cold branch and using
\begin{equation}
  \dot\xi_f=\frac{\dot T_{h,f}}{T_v}
  -\xi_f\frac{\dot T_v}{T_v}\,,
\end{equation}
one obtains Eq.~(\ref{eq:dxi_derivation}) after changing variables from $t$ to
$T_v$, $d\xi_f/dT_v=\dot\xi_f/\dot T_v$:
\begin{equation}
    \frac{d\xi_f}{dT_v} = \frac{1}{T_v\,\dfrac{d\rho_h}{dT_h}}
    \left[-\xi_f\,\frac{d\rho_h}{dT_h}
    + \frac{4H\,\sigma_h\rho_h - j_h}
    {4H\,\sigma\rho - 4H\,\sigma_h\rho_h - j_v}
    \,\frac{d\rho_v}{dT_v}\right],
    \label{eq:dxi_derivation}
\end{equation}
This is the standard two-sector temperature-ratio
equation~\cite{Li:2025nja,Feng:2024pab}, now identified with the exterior
temperature that enters the nucleation rate.

The reheated true-vacuum branch is obtained by adding the latent-heat source to
the hidden-sector equation,
\begin{align}
    \frac{d\rho_v}{dt} + 3H(\rho_v + p_v) &= j_v\,,
    \label{eq:drho_vR_dt} \\
    \frac{d\rho_{h,t}}{dt} + 3H(\rho_{h,t} + p_{h,t})
    &= j_h+\dot{\mathcal F}\,\Delta V\,.
    \label{eq:drho_hR_dt}
\end{align}
Thus
\begin{equation}
    \dot T_{h,t}
    = \frac{j_h+\dot{\mathcal F}\Delta V-4H\sigma_h\rho_h}
    {d\rho_h/dT_h}\,.
    \label{eq:Tdot_ht_app}
\end{equation}
The visible-sector clock remains Eq.~(\ref{eq:Tdot_v_app}), with the Hubble
rate evaluated from the volume-averaged energy density below.  Equivalently, the
latent heat modifies the hidden-temperature numerator but does not appear as a
direct heat source in the visible equation.  With
$T_{h,t}=\xi_tT_v$ and
\begin{equation}
  \dot\xi_t=\frac{\dot T_{h,t}}{T_v}
  -\xi_t\frac{\dot T_v}{T_v}\,,
\end{equation}
one obtains
\begin{equation}
    \frac{d\xi_t}{dT_v} = \frac{1}{T_v\,\dfrac{d\rho_h}{dT_h}}
    \left[-\xi_t\,\frac{d\rho_h}{dT_h}
    + \frac{4H\,\sigma_h\rho_h - j_h - \dot{\mathcal F}\,\Delta V}
    {4H\,\sigma\rho - 4H\,\sigma_h\rho_h - j_v}
    \,\frac{d\rho_v}{dT_v}\right].
    \label{eq:dxit_derivation}
\end{equation}
The visible temperature is used as the monotonic clock.  Since the released
latent heat is deposited in the hidden sector, not in the visible bath,
the appropriate time variable for bubble growth remains the visible-sector
clock.  For the cold branch $dT_{h,f}/dT_v\simeq \xi_f$ in the geometric
Jacobian, so
\begin{equation}
    \dot{\mathcal F}
    = \frac{d\mathcal F}{dT_{h,f}}\,\dot T_{h,f}
    \simeq -H\,T_{h,f}\,\frac{d\mathcal F}{dT_{h,f}}\,,
\end{equation}
where $d\mathcal F/dT_{h,f}<0$ as the universe cools.  Therefore
\begin{equation}
    \dot{\mathcal F}\,\Delta V
    = -H\,T_{h,f}\,\frac{d\mathcal F}{dT_{h,f}}\,\Delta V > 0\,,
    \label{eq:latent_source_current}
\end{equation}
where the derivative is evaluated on the cold branch.  Equations
(\ref{eq:dxi_derivation}) and (\ref{eq:dxit_derivation}) are the two
temperature-ratio equations used in Section~\ref{sec:boltzmann_review}.

The Hubble rate is common to both branches and is sourced by the
volume-averaged total energy density:
\begin{equation}
    H(T_v) = \sqrt{\frac{1}{3\Mpl^2}\left(
    \rho_v +(1-\mathcal{F})\rho_{h,f}+\mathcal{F}\rho_{h,t}
    + (1-\mathcal{F})\,\Delta V\right)}\,,
    \qquad 0 \le \mathcal{F} \le 1\,,
    \label{eq:Hubble}
\end{equation}
with $\rho_{h,f}$ evaluated at $T_{h,f}=\xi_fT_v$ and $\rho_{h,t}$ at
$T_{h,t}=\xi_tT_v$.  This is the appendix form of
Eq.~(\ref{eq:Hubble_3sector}).  The converted fraction $\mathcal F$ is obtained
from the moment hierarchy below, so the vacuum contribution drains
continuously as the transition completes.

\subsection{Derivation of the nucleation-moment hierarchy}
\label{sec:appendix_moment_derivation}

The false-vacuum fraction is $P_f=1-\mathcal{F}=e^{-I}$, where $I$ is the
expected true-vacuum volume nucleated per unit comoving volume.  Nucleation is
controlled by the cold exterior temperature $T_{h,f}$, so the bounce rate in
the following expressions is always $\Gamma(T_{h,f})$.  It is convenient to
parametrize time by the comoving distance
$h=\int dt/a$, which increases monotonically as the visible temperature cools.
A bubble nucleated at $h'$ and observed at $h$ has radius
$r=v_w(h-h')$.  Defining
$\ell(h)\equiv\Gamma(T_{h,f})/T_{h,f}^{\,4}$, the nucleation integral becomes
\begin{equation}
    I(h) = \frac{4\pi}{3}\,v_w^3\int_{0}^{h}\ell(h')\,[\,h-h'\,]^3\,dh'
    \equiv \frac{4\pi}{3}\,v_w^3\,U_3(h)\,,
    \label{eq:I_as_U3}
\end{equation}
where the cube encodes the bubble volume.  The nested form is inefficient to
evaluate during the ODE evolution.  We therefore introduce the moment family
\begin{equation}
    U_n(h) \equiv \int_{0}^{h}\ell(h')\,[\,h-h'\,]^n\,dh'\,,
    \qquad n = 0,1,2,3\,,
    \label{eq:Un_def}
\end{equation}
with $U_n=0$ at the onset.  Differentiating with respect to $h$ gives the closed
hierarchy
\begin{equation}
    \frac{dU_0}{dh} = \ell(h) = \frac{\Gamma(T_{h,f})}{T_{h,f}^{\,4}}\,,
    \qquad
    \frac{dU_n}{dh} = n\,U_{n-1}\,, \quad n = 1,2,3\,,
    \label{eq:moment_recursion_h}
\end{equation}
which replaces the nested integral by four first-order equations.  Since
$T_v$ remains monotonic through hidden-sector reheating, the system is evolved
in $T_v$.  The geometric Jacobian is
$dh/dT_v=-\xi_f/H$, giving the moment equations in
Eq.~(\ref{eq:ode_moments}).  The converted fraction and its derivative follow
algebraically,
\begin{equation}
    \mathcal{F} = 1 - e^{-I}\,,
    \qquad
    I = \frac{4\pi}{3}\,v_w^3\,U_3\,,
    \qquad
    \frac{d\mathcal{F}}{dT_{h,f}}
    = -(1-\mathcal{F})\,\frac{4\pi\,v_w^3\,U_2}{H}\,,
    \label{eq:F_closure}
\end{equation}
which supplies the latent-heat source in Eq.~(\ref{eq:latent_source_current}).

\subsection{Complete 11-ODE system}

We collect the explicit evolution equations for the state vector\\
$\mathbf{y} = (\xi_f,\,\xi_t,\, Y_{\tilde\chi},\, Y_{\gamma'},\,
Y_\chi,\, Y_q,\,Y_{\sigma},\, U_0,\, U_1,\, U_2,\, U_3)$.
The 11 components are two temperature ratios, five particle yields, and four
nucleation moments.

\paragraph{Temperature ratios $\xi_f(T_v)$ and $\xi_t(T_v)$ (2 equations).}
The cold exterior evolves according to
\begin{equation}
    \frac{d\xi_f}{dT_v} = \frac{1}{T_v\,\dfrac{d\rho_h}{dT_h}}
    \left[-\xi_f\,\frac{d\rho_h}{dT_h}
    + \frac{4H\,\sigma_h\rho_h - j_h}
    {4H\,\sigma\rho - 4H\,\sigma_h\rho_h - j_v}
    \,\frac{d\rho_v}{dT_v}\right],
    \label{eq:ode_xi}
\end{equation}
with $T_h=T_{h,f}=\xi_fT_v$.  The reheated true-vacuum interior obeys
\begin{equation}
    \frac{d\xi_t}{dT_v} = \frac{1}{T_v\,\dfrac{d\rho_h}{dT_h}}
    \left[-\xi_t\,\frac{d\rho_h}{dT_h}
    + \frac{4H\,\sigma_h\rho_h - j_h - \dot{\mathcal F}\,\Delta V}
    {4H\,\sigma\rho - 4H\,\sigma_h\rho_h - j_v}
    \,\frac{d\rho_v}{dT_v}\right],
    \label{eq:ode_xit}
\end{equation}
with $T_h=T_{h,t}=\xi_tT_v$ in the thermodynamic functions of the true branch.
The source is
$\dot{\mathcal F}\Delta V=-H\,T_{h,f}(d\mathcal F/dT_{h,f})\Delta V$.
The Hubble rate is the common volume-averaged rate of Eq.~(\ref{eq:Hubble}).

\paragraph{Nucleation moments $U_0$--$U_3$ (4 equations).}
The moments are evolved with the cold exterior temperature:
\begin{equation}
    \frac{dU_0}{dT_v} = -\frac{\xi_f\,\Gamma(T_{h,f})}
    {T_{h,f}^{\,4}\,H},
    \qquad
    \frac{dU_n}{dT_v} = -\frac{n\,\xi_f\,U_{n-1}}{H},
    \quad n = 1,2,3
    \label{eq:ode_moments}
\end{equation}
The factor $\xi_f/H$ is the exact comoving-distance Jacobian with the visible
temperature as clock.  It should not be replaced by $dT_{h,t}/dT_v$, which can
vanish during reheating even though time and bubble growth continue.  The
moments vanish at the critical temperature, and
$I=(4\pi/3)v_w^3U_3$, $\mathcal F=1-e^{-I}$.

\paragraph{Gaugino yield.}
\begin{equation}
    \frac{dY_{\tilde\chi}}{dT_v} = \mathcal{J}\left[
    \langle\sigma v\rangle_{\tilde\chi\tilde\chi\to\chi\chi}^{\rm eff}
    + \langle\sigma v\rangle_{\tilde\chi\tilde\chi\to q\bar q}^{\rm eff}
    \right](T_{h,f})\left(Y_{\tilde\chi}^{\rm eq}(T_{h,f})^2
    - Y_{\tilde\chi}^2\right),
    \label{eq:ode_gaug}
\end{equation}
where $Y_{\tilde\chi}^{\rm eq} = (45/(2\pi^4))\,(m_{\rm eff}/T_{h,f})^2
K_2(m_{\rm eff}/T_{h,f})\,\xi_f^3/h_{*,v}$ is the equilibrium yield
summed over both eigenstates. In practice the squark channel
contributes $\lesssim 0.01\%$ and can be neglected.
Inside the bubbles, the gaugino eigenstates decay instantaneously:
$\tilde\chi_+ \to \gamma'\chi$ ($\tau \sim 10^{-30}\;$s) and
$\tilde\chi_- \to q\bar q\chi$ (3-body, $\tau \sim 10^{-25}\;$s).

\paragraph{Dark photon yield.}
In the symmetric phase, the dark photon $\gamma'$ is massless and
in thermal equilibrium with the hidden sector bath. The only
coupling to the visible sector is through kinetic mixing
($\delta$), which mediates production via inverse decay
$ii \to \gamma'$ and decay $\gamma' \to ii$ (where $ii$ denotes
visible-sector fermion pairs). The yield equation is
\begin{equation}
    \frac{dY_{\gamma'}}{dT_v} = \mathcal{J}\,\left[
    \frac{\Mpl^2\,\langle\sigma v\rangle_{ii\to\gamma'}}{s^2}
    - \frac{\Gamma_{\gamma'\to ii}}{s}\,Y_{\gamma'}\,\Mpl
    \right],
    \label{eq:ode_gp}
\end{equation}
where $\langle\sigma v\rangle_{ii\to\gamma'}$ is the thermally
averaged production cross section from the visible bath and the
partial decay width
$\Gamma_{\gamma'\to ii} \propto \delta^2\,\alpha_{\rm em}\,m_{\gamma'}$
(Appendix~\ref{sec:appendix_xsec}).
Both terms are proportional to $\delta^2$ and are negligible for
$\delta \lesssim 10^{-7}$; the dark photon yield then tracks its
thermal equilibrium value until the FOPT.
Inside the bubbles, $\gamma'$ acquires mass
$m_{\gamma'} = g_x v_X$ and decays instantly to dark quarks
($\gamma' \to q\bar q$, $\tau \sim 10^{-30}\;$s).

\paragraph{$\sigma$ yield.}
The lighter Higgs scalar $\sigma$ pair-annihilates through two
channels. The dominant channel
$\sigma\sigma \to \gamma'\gamma'$ proceeds via the quartic
gauge vertex $\frac{1}{2}g_x^2\,\sigma^2\,A_\mu A^\mu$ (from
$|D_\mu\Phi|^2$) plus $t/u$-channel exchange, with cross section
given in Eq.~(\ref{eq:sv_sigD_gg}).
The subdominant channel $\sigma\sigma \to q\bar q$ proceeds
via $s$-channel $\gamma'$ exchange:
\begin{align}
    \sigma v\big|_{\sigma\sigma \to q\bar q}
    = \frac{g_x^4 q_X^2}{4\pi}\,
    \frac{s}{(s - m_A^2)^2 + m_A^2\Gamma_{\gamma'}^2}\,
    \sqrt{1 - \frac{4m_q^2}{s}}\,,
\end{align}
where $\Gamma_{\gamma'} = \alpha_X q_X^2 m_{\gamma'}/3$ is the
dark photon decay width. In the symmetric phase ($m_A = 0$), the
$s$-channel propagator is unsuppressed but the overall rate is
subdominant by a factor of $q_X^2$ relative to the gauge-boson
final state. The full yield equation is
\begin{equation}
    \frac{dY_{\sigma}}{dT_v} = \mathcal{J}\left[
    \langle\sigma v\rangle_{\sigma\sigma\to\gamma'\gamma'}
    + \langle\sigma v\rangle_{\sigma\sigma\to q\bar q}
    \right](T_{h,f})
    \left(Y_{\sigma}^{\rm eq}(T_{h,f})^2
    - Y_{\sigma}^2\right),
    \label{eq:ode_sigD}
\end{equation}
where $Y_{\sigma}^{\rm eq}$ is the scalar equilibrium yield
at $T_{h,f}$.
Inside the bubbles, $\sigma$ decays instantly via
$\sigma \to \gamma'\chi$ ($\tau \sim 10^{-30}\;$s).

\paragraph{Hidden Higgs $\chi$.}
In the symmetric phase, $\chi$ maintains equilibrium through
annihilation $\chi\chi \to \gamma'\gamma'$ via $t/u$-channel
$\chi$ exchange with the gauge vertex from $|D_\mu\Phi|^2$.
For massless $\gamma'$ and $m_\chi \ll T_h$, the $s$-wave cross
section is
\begin{align}
    \sigma v\big|_{\chi\chi \to \gamma'\gamma'}
    = \frac{g_x^4}{16\pi\,m_\chi^2}\,.
\end{align}
The yield equation is
\begin{equation}
    \frac{dY_\chi}{dT_v} = \mathcal{J}\,
    \langle\sigma v\rangle_{\chi\chi\to\gamma'\gamma'}(T_{h,f})
    \left(Y_\chi^{\rm eq}(T_{h,f})^2 - Y_\chi^2\right).
    \label{eq:dYchi_dTv}
\end{equation}
On the flat direction ($\lambda_h = 0$), there are no
number-changing processes beyond the $\gamma'$ channel above,
which becomes kinematically forbidden after the FOPT because
$m_{\gamma'} = g_x v_X \approx 8.5\times 10^6\;\gev \gg m_\chi
\sim g_x^2 m_0/(4\pi) \sim 4\times 10^5\;\gev$.
The hidden Higgs therefore freezes out at $\Tp$ with
$Y_\chi \sim 10^{-4}$ (Fig.~\ref{fig:yields}).
After the FOPT, the yield evolves only through
gravity-mediated decay ($\chi \to$ SM particles, with branching
ratio to $q\bar q$ via the Yukawa coupling):
\begin{equation}
    \frac{dY_\chi}{dT_v} = -\frac{\Gamma_\chi^{\rm tot}}{H\,T_v}\,Y_\chi\,,
    \label{eq:dYchi_decay}
\end{equation}
where $\Gamma_\chi^{\rm tot} \sim m_\chi^3/\Mpl^2$
(Appendix~\ref{sec:appendix_xsec}),
giving
 $\tau_\chi \sim 10^{-5}\;$s. 
This decay creates the
matter-dominated epoch and entropy dilution $D \sim 600$
discussed in Section~\ref{sec:relic_density}.

\paragraph{Dark quark $q$.}

The dark quark yield obeys the standard Boltzmann equation
with three annihilation channels
(Appendix~\ref{sec:appendix_dq_xsec}):
\begin{equation}
    \frac{dY_q}{dT_v} = \mathcal{J}\left[
    \langle\sigma v\rangle_{q\bar q\to\gamma'\gamma'}
    + \langle\sigma v\rangle_{q\bar q\to\tilde\lambda\tilde\lambda}
    + \langle\sigma v\rangle_{q\bar q\to\tilde H\tilde{\bar H}}
    \right](T_{h,f})
    \left(Y_q^{\rm eq}(T_{h,f})^2 - Y_q^2\right),
    \label{eq:dYq_dTv}
\end{equation}
where $Y_q^{\rm eq} = (45\zeta(3)/\pi^4)\,g_q\,\xi_f^3/h_{*,v}$
with $g_q = 4$ (Dirac fermion).
In practice, the combined interaction rate
$\Gamma_{\rm tot}/H \approx 3\times 10^8$ at $T_{h,f} = \Tp$
(Eq.~(\ref{eq:Gamma_over_H})) keeps the dark quark deep in
chemical equilibrium throughout the symmetric phase, so
$Y_q$ tracks $Y_q^{\rm eq}$ to high precision.
After the FOPT, all three channels shut off simultaneously as the
gauge-sector species acquire masses $\sim g_x v_X \gg T_p$,
and $Y_q$ freezes instantaneously at the relativistic
equilibrium value $Y_q^{(0)} \approx 2.1$--$2.3\times 10^{-4}$
(Table~\ref{tab:cosmo_constraints}).
The yield is subsequently modified only by the entropy dilution
from $\chi$ decay (Section~\ref{sec:relic_density}).

In summary, the yield equations retain their standard form, but their
equilibrium quantities are evaluated on the cold branch
$T_{h,f}=\xi_fT_v$ until the phase transition shuts off the corresponding
channels.  The vacuum-energy release enters the yields indirectly through the
temperature history and through the rapid decays inside the bubbles.  The cross
sections appearing in the yield equations are collected in
Appendix~\ref{sec:appendix_xsec}.

\end{document}